\documentclass[%
 reprint,
 amsmath,amssymb,
  pra,
]{revtex4-1}

\usepackage{graphicx}
\usepackage{dcolumn}
\usepackage{bm}

\begin{document}


\title{Phase Estimation with Non-Unitary Interferometers: Information as a Metric}

\author{Thomas B. Bahder}
\affiliation{%
Aviation and Missile Research, 
      Development, and Engineering Center, \\    
US Army RDECOM, 
Redstone Arsenal, AL 35898, 
U.S.A.}%

\date{\today}

\begin{abstract}
Determining the phase in one arm of a quantum interferometer is discussed taking into account the three non-ideal aspects in real experiments: non-deterministic state preparation, non-unitary state evolution due to losses during state propagation, and imperfect state detection.   A general expression is written for the probability of a measurement outcome taking into account these three non-ideal aspects.  As an example of applying the formalism, the classical Fisher information and fidelity (Shannon mutual information between phase and measurements) are computed for few-photon Fock and N00N  states input into a lossy Mach-Zehnder interferometer.  These three non-ideal aspects lead to qualitative differences in phase estimation, such as a decrease in fidelity and Fisher information that depends on the true value of the phase.      
\end{abstract}

\maketitle


\section{\label{Intro}Introduction}

Optical interferometers~\cite{Hariharan2003} and matter wave interferometers~\cite{Cronin2009} have been of great interest because of their practical applications in metrology.  Interferometers have been used to measure such diverse quantities as electric, magnetic, and gravitational fields, gravitational waves~\cite{Cronin2009,Thorne1980,Caves1981}, and there are plans to use them to test the theory of general relativity~\cite{Dimopoulos2008}.  Classical optical interferometers~\cite{Lefevre1983} have been routinely used for sensing rotation in gyroscopic applications based on the Sagnac effect~\mbox{\cite{Sagnac1913a,Sagnac1913b,Sagnac1914,Post1967,Chen2008}} and experiments with Sagnac interferometers have been done with single-photons~\cite{Bertocchi2006}, with Bose-Einstein condensates(BEC)~\cite{Gupta2005,Wang2005,Tolstikhin2005}, and schemes using entangled particles have been proposed that are capable of Heisenberg limited precision measurements that scale as $1/N$, where $N$ is the number of particles~\cite{Cooper2010}.    

On a more fundamental level, there is interest in interferometers because they are a vehicle to study the limits of precision of quantum measurements~\cite{Godun2001,Giovannetti2006,Berry2009}.   Perhaps the simplest generic measurement problem consists of determining the relative phase shift between two arms of an interferometer from measurements made at the output ports of the interferometer~\cite{Combes2005,Nagata2007,Durkin2007,Pezze2008,Dorner2009,Cable2010}. This phase shift may be related to a classical external field incident on a phase shifter in one arm of the interferometer, in which case the interferometer can be used as a sensor of the field~\cite{Bahder2006}.  The determination of the phase shift is a specific example of the more general problem of parameter estimation, whose goal is to determine one or more parameters from measurements~\cite{Cramer1958,Helstrom1967,Helstrom1976,Holevo1982,Braunstein1994,Braunstein1996,Barndorff-Nielsen2000,Barndorff-Nielsen2003}.  

Recently, there have been experimental demonstrations using entangled states to estimate the phase shift in one arm of a Mach-Zehnder interferometer~\mbox{\cite{Walther2004,Mitchell2004,Nagata2007,Okamoto2008}}.  Even more recently, the effect of losses on phase determination was studied experimentally~\cite{Kacprowicz2010}.  In real experiments, there are three non-ideal elements of the interferometer system:  state preparation~\cite{Mitchell2004,Thomas-Peter2009}, photon losses in the interferometer~\cite{Kacprowicz2010} and non-ideal photon-number detection~\cite{Nagata2007,Okamoto2008}.  Phase estimation has been theoretically investigated by separately taking into account non-deterministic state preparation~\mbox{\cite{Helstrom1967,Helstrom1976,Holevo1982,Braunstein1994,Braunstein1996,Barndorff-Nielsen2000,Barndorff-Nielsen2003}}, photon losses in the interferometer itself~\cite{Kim1998,Durkin2004,Rubin2007,Gilbert2008,Dorner2009,Demkowicz-Dobrzanski2009,Cable2010,Ono2010} and photon-number counting efficiency~\cite{Okamoto2008,D'Ariano2000,Cable2010}.

In this work, I write down a formalism that simultaneously takes into account, in a unified way, all three of the non-ideal elements in experiments:  non-deterministic state preparation, propagation through a lossy interferometer, and imperfect state detection.  Non-deterministic state preparation must be described by a density matrix, rather than by a pure state, thereby allowing for the finite probability of creating states other than intended.  When the optical state is created, it enters the interferometer, where propagation may be non-ideal because photon absorption and scattering can occur. Finally, when the optical state leaves the interferometer, it enters the detection system, which may also be non-ideal:  the state registered by the detection system may not be the true state that entered the detection system. 

Much of the previous work was focused on determining the optimum measurements for determining phase and hence the quantum Fisher information was of primary interest, because it gives a bound on the variance of the phase associated with the optimum measurement~\mbox{\cite{Helstrom1967,Helstrom1976,Holevo1982,Braunstein1994,Braunstein1996,Barndorff-Nielsen2000,D'Ariano2000,Monras2006,Olivares2009,Gaiba2009}}.   In contrast, in this work I look at the information gain from specific, simple, photon-number counting measurements that can be easily implemented in the laboratory, and hence the classical Fisher information is the quantity of interest because it depends on the particular measurement that is performed.   

In Section~\ref{TheoryBackgroudSection}, I briefly review the theory of phase determination based on parameter estimation (Fisher information)  and on fidelity (Shannon mutual information between measurements and phase).  In Section~\ref{Non-IdealSystem}, I introduce an example of a non-ideal interferometer system, where state evolution is non-unitary.  I write a statistical expression for the probability of measurement outcomes that takes into account the three non-ideal components of the interferometer system described above.  I use this probability in the classical Fisher information in Eq.~(\ref{ClassicalFisherInformation}) and in the fidelity in Eq.~(\ref{ShannonMutualInformation}) to analyze the determination of phase in a non-ideal interferometer system.  As simple examples of the  formalism, in sub-sections of Section~\ref{Non-IdealSystem}, I look at few photon examples of non-deterministic state preparation, propagation through a non-unitary (lossy) interferometer, and imperfect state detection.  Finally, in Section~\ref{Conclusion}, I make some concluding remarks.    My goal is to look at examples of few-photon states that can be implemented experimentally, with the hope that the examples and method described here can be helpful for analyzing real experiments.  Furthermore, in this work, I restrict myself to the simple case of non-adaptive measurements~\cite{Higgins2009}, where the measurement is fixed before phase estimation.

\section{\label{TheoryBackgroudSection}Theoretical background}

The accuracy of estimating a (single) one-dimensional parameter, $\phi$, is described in terms of the classical Cramer-Rao bound~\cite{Cover2006}, which gives a lower bound on the variance $(\delta\phi)^2$ of an unbiased estimator of the parameter $\phi$:
\begin{equation}
\left( {\delta \phi  } \right)^2  \ge \frac{1}{ F_{cl}(\phi;M )}
\label{Cramer-RaoBound}
\end{equation}
where $F_{cl}(\phi;M )$ is the classical Fisher information given by~\cite{Cramer1958,Cover2006}
\begin{equation}
F_{cl}(\phi;M ) = \sum\limits_\xi {\frac{1}{{P(\xi|\phi, \rho )}}\,\left[ {\frac{{\partial P(\xi|\phi, \rho )}}{{\partial \phi }}} \right]^2 } 
\label{ClassicalFisherInformation}
\end{equation}
The classical Fisher information is described in terms of the  conditional probability distribution,  $P(\xi|\phi,\rho)$, for measurement outcome, $\xi$, which can take one or more continuous values, or, one or more discreet values. If $\xi$ takes continuous values, the sum over $\xi$ in Eq.~(\ref{ClassicalFisherInformation}) is an integral.  For the case of quantum measurements, these probabilities are given by 
\begin{equation}
P(\xi |\phi, \rho ) = {\rm{tr}}\left( {\hat \rho_\phi \,\hat \Pi_o  \left( \xi \right)} \right) = {\rm{tr}}\left( {\hat \rho_o \,\hat{\Pi}_\phi  \left( \xi \right)} \right)
\label{MeasurementProbabilityDensitymatrix}
\end{equation}
where the state is specified  by the  Schr\"odinger picture density matrix, $\hat{\rho}_\phi$, and the measurements by the positive-operator valued  measure (POVM) in the Schr\"odinger picture, $\hat \Pi_o ( \xi )$. 
The POVM are set a of non-negative Hermitian operators, $M= \{ \hat \Pi (\xi) \}$, representing a given physical measurement and so the expectation value of each operator $\hat \Pi_o (\xi)$ is non-negative, and satisfies $\sum\limits_\xi {\hat \Pi_o (\xi)}  = \hat{I}$, where $\hat{I}$ is the identity operator.  Alternatively,  in the Heisenberg picture,  the state is given by the density matrix, $\hat{\rho}_o$, and probabilities of measurements by POVM, $\hat \Pi_\phi ( \xi )$.  Operators in the Schrodinger and Heisenberg pictures,  $\hat{O}_S$ and $\hat{O}_H(\phi)$, respectively are related  by $\hat{O}_H(\phi) = \hat{U}^\dagger(\phi) \, \hat{O}_S \, \hat{U}(\phi)$, and $\hat{U}(\phi)=\exp(-i \phi \hat{h})$, where $\hat{h}$ is the infinitesimal displacement operator for the parameter $\phi$, satisfying  
\begin{equation}
i \frac{\partial }{\partial  \phi} \left| \psi (\phi)  \right\rangle = \hat{h} \, \left| \psi(\phi)  \right\rangle 
\label{displacementOp}
\end{equation}

The quantum Fisher information, $F_{Q}(\phi)$ is obtained by maximizing the classical Fisher information, $F_{cl}(\phi;M )$, over all possible measurements $M$, at a given value of $\phi$.  Braunstein and Caves~\cite{Braunstein1994,Braunstein1996} have shown that an improved lower bound is possible for the variance, $\left( {\delta \phi  } \right)^2$, in terms of the quantum Fisher information: 
\begin{equation}
\left( {\delta \phi  } \right)^2  \ge \frac{1}{{F_{cl}(\phi;M )}} \ge \frac{1}{{F_{Q}(\phi)}}
\label{QuantumInequality}
\end{equation}
where $F_{Q}(\phi)$ is independent of the measurement $M$.  The quantum Fisher information is defined by 
\begin{equation}
F_Q \left( \phi \right) = \rm{tr} \left[ \hat{\rho}_\phi  \hat{\Lambda}_{\phi}^2  \right]
\label{QuantumFisherInfo}
\end{equation}
where the Hermitian operator, $\Lambda_{\phi}$, is the symmetric logarithmic derivative (S.L.D.), defined implicitly by 
\begin{equation}
\frac{\partial \hat{\rho}_\phi  }{\partial \phi } = \frac{1}{2}\left[ \hat{\Lambda}_{\phi} \, \hat{\rho}_\phi  + \hat{\rho}_\phi  \hat{\Lambda}_{\phi} \right]
\label{DerivRho}
\end{equation}
The right side and left side of the inequality in Eq.~(\ref{QuantumInequality}) is sometimes called the {\it quantum} Cramer-Rao bound, see also the work by Helstrom~\cite{Helstrom1967,Helstrom1976} and Holevo~\cite{Holevo1982}, and discussion by Barndorff-Nielsen et al.~\cite{Barndorff-Nielsen2000,Barndorff-Nielsen2003}.  The expression in Eq.~(\ref{QuantumInequality}) provides a bound on the variance of an unbiased estimator for an optimum measurement. However, the theory does not give a procedure for determining the optimum measurement. 
For one-dimensional parameter estimation and for simple (non-adaptive) measurements, Barndorff-Nielsen and Gill have shown that in general the optimum measurement $M$ will depend on the parameter $\phi$, which is unknown prior to estimation~\cite{Barndorff-Nielsen2000}.  Consequently, Barndorff-Nielsen and Gill have proposed a two-stage adaptive measurement procedure that will give 
\begin{equation}
F_{cl}(\phi; M )  = F_Q \left( \phi \right)
\label{ClassicalQuantumFisher}
\end{equation}
for optimum measurement $M$ for all $\phi$.     

For the case of a pure state, $\left| {\psi _o } \right\rangle$, where the density matrix is $\rho _o  = \left| {\psi _o } \right\rangle \,\left\langle {\psi _o } \right|$, and where the path is generated by a unitary transformation, $\hat{U}(\phi)$, the quantum Fisher information, $F_Q( \phi)$,  does not depend on $\phi$~\cite{Braunstein1996,Olivares2009,Gaiba2009}, and is given by the fluctuations of the generator $\hat{h}$ by $F_Q \left( \phi \right) = 4 ( \Delta h)^2$.   Furthermore, Hofman has shown that for pure states having a path symmetry~\cite{Hofmann2009}, the quantum Cramer-Rao bound in Eq.(\ref{QuantumInequality}) can be achieved at any value of phase $\phi$. 
The condition for optimal measurements for the case of pure states has also been investigated~\cite{Durkin2010}.

In Section~\ref{Non-IdealSystem} below, I consider the case of a lossy Mach-Zehnder interferometer, where the state evolution is effectively non-unitary, thereby leading to a classical Fisher information that depends on the true value of the phase $\phi$.

The above discussion of parameter estimation is based on classical and quantum Fisher informations, which are local descriptions of phase estimation, because they depend on the true value of $\phi$.   Complementary to the above local descriptions, is a global description given by the fidelity~\cite{Bahder2006}:
\begin{eqnarray}
H(M) & = & \sum_{\xi}\int_{-\pi}^{+\pi}d\,\phi\,\,P(\xi|\phi, \rho
) \,  p(\phi) \,\, \times \,    \nonumber  \\
  &  & \log_{2}\left[  \frac{ P(\xi|\phi,\rho)\,}{\int_{-\pi}^{+\pi}%
\,\,\,P(\xi|\phi^{\prime},\rho)  \, p(\phi^\prime) \,\,d\,\phi^{\prime}}\right].
\label{ShannonMutualInformation}%
\end{eqnarray}
%
where $P(\xi|\phi,\rho)$ is given in Eq.~(\ref{MeasurementProbabilityDensitymatrix}). 
The fidelity,  $H(M)$, is the Shannon mutual information~\cite{Shannon1948,Cover2006} between the measurement $M$ and the unknown parameter $\phi$.   The fidelity, $H(M)$, gives the average amount of information (in bits) about the parameter $\phi$ that can be obtained from the measurement $M$ for one use (one measurement cycle) of the interferometer.  The fidelity does not depend on $\phi$ because it is an average over all possible phases $\phi$ and over all probabilities of measurement outcomes for a given POVM $M$. (For an alternative discussion of local versus global phase estimation, see~Ref.\cite{Durkin2007}.) The fidelity also depends on the prior information about the parameter $\phi$ through the prior probability distribution $p(\phi)$.   Consequently, the fidelity characterizes the quality of the interferometer system as a whole, in terms of mutual information between the measurement $M$ and the parameter $\phi$.  Note that the fidelity depends on the input state density matrix, $\hat{\rho}$, and the measurement $M$, and therefore can be used to optimize the system with respect to the input state and measurement.  The fidelity is a measure of the information that flows from the phase $\phi$ to the measurements, which is analogous to a communication problem where Alice sends messages to Bob.  In the case of the measurement problem, quantum fluctuations in the initial state, in the channel (interferometer), and the type of measurement,  determine the amount of information that is obtained about the parameter $\phi$ from the measurements. The fidelity has been applied to compare the use of Fock states and N00N states when no prior information is present about the phase~\cite{Bahder2006} and when there is significant prior information about the phase~\cite{Bahder2007}.

The complimentary measures of fidelity and Fisher information may be contrasted as follows.  Assume that I want to shop to purchase the best measurement device to determine the unknown parameter $\phi$. If I do not know the true value of the parameter $\phi$,  I would compare the overall performance specifications of several devices and I would purchase the device with the best over-all specifications for measuring $\phi$. The fidelity, $H(M)$, is the over-all specification for the quality of the device, so I would purchase the device with the largest fidelity.  After I have purchased the device, I want to use it to determine a specific value of the  parameter $\phi$ based on several measurements (data).  This involves parameter estimation, which requires the use of Fisher information, and depends on the true value of the parameter $\phi$.

Historically, the variance, $(\delta \phi)^2$, of the estimated parameter $\phi$ has been discussed in terms of the standard quantum limit, $\delta \phi_{SL}=$ $1/\sqrt{N}$, and the Heisenberg limit\cite{Caves1981,Ou1997,Giovannetti2004,Giovannetti2006}, $\delta\phi_{HL}=$ $1/N$, where $N$ is the number of particles or quanta that enter the interferometer during each measurement cycle.  The value $\delta \phi$ is presumably the width of some probability distribution, $p(\phi | \xi,\rho)$, such as the distribution calculated from Bayes' rule, see Eq.~(\ref{PhaseProbabilityDistribution}).  Detailed calculation of $p(\phi | \xi,\rho)$ for a number of input states shows that these  distributions  have multiple peaks~\cite{Bahder2006}.  Consequently, rather than using the widths of these distributions as a metric for determining $\phi$, I use the  information measures, Fisher information and fidelity, which naturally handle distributions with multiple peaks.

\begin{figure}[t]   
\includegraphics[width=3.3in]{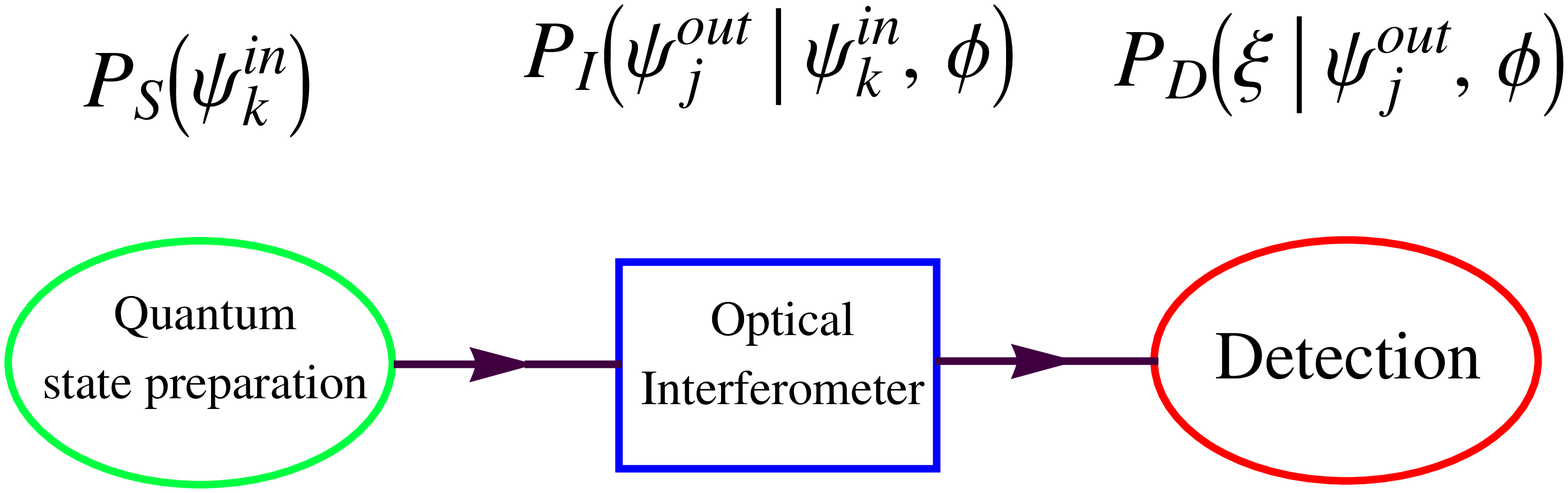}
\caption{\label{fig:InterferometerSystem}(Color) Interferometer system shown with three components: state preparation, interferometer, and detection system.}
\end{figure}

\section{\label{Non-IdealSystem}Non-Ideal Optical System}
As described in the introduction, an interferometer system can be divided into three parts: state creation, state evolution through the optical interferometer, and detection of the output state.  In a real experiment,  each of these three parts can be non-ideal, see Fig.~\ref{fig:InterferometerSystem}.  For example, I may want to create a quantum state $|\psi^{in}\rangle$ as input into the interferometer.  However, instead,  the resulting state may be a mixture of states, each with some probability, $P_S (\psi _k^{in} )$, for $k=1,2,\cdots$.  Such a quantum state is described by the density matrix $\hat{\rho}$:
\begin{equation}
\hat{\rho}  = \sum\limits_k {\,\;P_S (\psi _k^{in} )} \;\left| {\psi _k^{in} } \right\rangle \left\langle {\psi _k^{in} } \right|
\label{DensityMatrix}
\end{equation}
The state  $\hat{\rho}$ is then input into the interferometer, where there may be absorption and scattering of photons.  For example, a two-photon state may enter the interferometer and a one-photon state may exit the interferometer, because one photon was absorbed inside the interferometer.  Alternatively, a two-photon state may enter the interferometer and a three-photon state may exit the interferometer, due to light scattering into the interferometer from the environment.  I can describe these processes generally by a transfer matrix,  ${P_I (\psi _j^{out} |\psi _k^{in} ,\phi )}$, which gives the conditional probability for state $\left| {\psi _j^{out} } \right\rangle $ to exit the interferometer given that state  $\left| {\psi _k^{in} } \right\rangle $ entered the interferometer. The transfer matrix,  ${P_I (\psi _j^{out} |\psi _k^{in} ,\phi )}$, is general enough to  describe  non-unitary propagation of the quantum state through the interferometer, and so can take into account losses and scattering.  Note that the transfer matrix may depend on the state of the interferometer, which I specify here by single parameter $\phi$.  Finally, the detection of the quantum state that leaves the interferometer can be non-ideal.  For example, the detection system may register a measurement $\xi$, when state $\left| {\psi _i^{out}} \right\rangle$ enters the detection system, whereas the true state that entered the detection system was $ \left| {\psi _j^{out} } \right\rangle$.  I can represent such an imperfect detection system by the conditional probability $P_D (\xi |\psi _j^{out} ,\phi )$, which gives the probability for making a measurement $\xi$ when state  $\psi _j^{out}$ entered the detection system.  Note that in general this probability may or may not depend on $\phi$, a parameter describing the state of the interferometer.  For a non-ideal interferometer system, the probability of obtaining a measurement $\xi$ is  given by~\cite{Jaynes2003}
\begin{equation}
\small
P(\xi |\phi ) = \sum\limits_j {P_D } (\xi |\psi _j^{out} ,\phi )\;\sum\limits_k {\,P_I (\psi _j^{out} |\psi _k^{in} ,\phi )\;P_S (\psi _k^{in} )} 
\label{MeasurementProbability}
\end{equation}
where we must have each of the three probabilities sum to unity:
\begin{equation}
\sum\limits_k {\,P_S (\psi _k^{in} )}  = 1
\label{StatepreparationSumUnity}%
\end{equation}
\begin{equation}
\sum\limits_j {\,P_I (\psi _j^{out} |\psi _k^{in} ,\phi )}  = 1
\label{InterferometerSumUnity}%
\end{equation}
\begin{equation}
\sum\limits_\xi {P_D (\xi |\psi _j^{out} ,\phi )}  = 1 
\label{DetectionSumUnity}%
\end{equation}

Equation~(\ref{MeasurementProbability}) is a general statistical relation for the probability of obtaining a measurement outcome $\xi$ for given phase shift $\phi$, taking into account the three non-ideal aspects of interferometer systems.  Note that Eq.~(\ref{MeasurementProbability}) is sufficiently general that it can be applied to the case where states are represented by density matrices.  In this case, in Eq.~(\ref{MeasurementProbability})  we can make the replacements $\psi _k^{in} \rightarrow \rho_k^{in} $ and  $\psi _j^{out} \rightarrow \rho_j^{out}$, where $\rho_k^{in}$ and $\rho_j^{out}$ are a set of input and output density matrices labeled by integers $j,k=1,2, \cdots$.  In order to compute the probabilities of measurement outcomes, $P(\xi |\phi )$, Eq.~(\ref{MeasurementProbability})  must be augmented by a detailed model of input and output states. I give several examples of applying Eq.~(\ref{MeasurementProbability}) in the sections that follow.  

The probability of measurement outcome, given by Eq.~(\ref{MeasurementProbability}), enters into the Fisher information and into the Shannon mutual information, in Eq.~(\ref{ClassicalFisherInformation}) and Eq.~(\ref{ShannonMutualInformation}), respectively.  In the next three subsections, A, B, and C, I give examples of the effects of non-deterministic state preparation, state evolution through an interferometer when absorption is present, and imperfect output state detection, respectively, using Fisher and Shannon mutual informations as metrics of performance of the interferometer.

\subsection{\label{Non-DeterministicStatepreparation}Non-Deterministic State Preparation}
Consider an optical interferometer system that has non-deterministic state preparation, but has no losses in the interferometer  and  has perfect state detection.   When I try to prepare a certain quantum state for input into the interferometer, there is always a non-zero probability that another state than intended will be prepared. This non-deterministic state preparation is expressed by a density matrix for the input state, which assigns probabilities for creating various quantum states, see Eq.~(\ref{DensityMatrix}).  Since state detection is assumed perfect, $P_D (\xi |\psi^{out} ,\phi )=1$ when the measurement $\xi$ corresponds to the true state that entered the detection system, $\psi^{out}$, and otherwise  $P_D (\xi |\psi^{out} ,\phi )=0$.   

A general interferometer with no losses is characterized by a unitary scattering matrix, $S_{ij}(\phi)$, that connects the  $N_p$ input-mode field operators,  $\hat{\alpha}_{i}$,  to the $N_p$ output field operators, $\hat{\beta}_{i}$:
\begin{equation}
\hat{\beta}_{i}=\sum_{j=1}^{N_p}S_{ij}(\phi)\,\hat{\alpha}_{j} = \hat U^\dag(\phi)  \hat \alpha _i \hat U(\phi)
\label{InputOutputDef}%
\end{equation}
where $\hat{U}(\phi)$ is a unitary evolution operator, $i,j=1,2,\cdots,N_p$, and  $\phi$ is one or more parameters (e.g., phase shift) that describe the state of the  interferometer.  

For simplicity, I consider a  Mach-Zehnder interferometer, with no losses,  with input ports labeled, ``a" and ``b", and output ports, ``c" and ``d", having a scattering matrix
\begin{equation}
S_{ij}(\phi)= \frac{1}{2}  \left(
\begin{array}{cc}
 - i \left(1+e^{i \phi
   }\right) & 
   \left(-1+e^{i \phi }\right) \\
  \left(-1+e^{i \phi
   }\right) &  i
   \left(1+e^{i \phi }\right)
\end{array}
\right)
\label{Smatrix2x2}
\end{equation}
where $\hat{\alpha}_{i} = \{ \hat{a}, \hat{b} \}$ and $\hat{\beta}_{i} = \{ \hat{c}, \hat{d} \}$.

The probabilities, $P_I (\psi _j^{out} |\psi _k^{in} ,\phi )$,  that relate the input state $\psi _k^{in}$ to the output state $\psi _j^{out}$ of the interferometer are given in terms of the projection operators $\hat \Pi _\phi  \left( {n_c ,n_d } \right)$:
\begin{equation}
P_I (\psi _j^{out} |\psi _k^{in} ,\phi ) = \left\langle {\psi _k^{in} } \right|\,\hat \Pi _\phi  \left( {n_c ,n_d } \right)\,\left| {\psi _k^{in} } \right\rangle 
\label{MeasurementOps}%
\end{equation}
where  the output state $\psi _j^{out}$ is specified by two integers, $ \left\{ n_c ,n_d  \right\}$, giving the photon numbers output in ports ``c" and ``d".   In terms of the unitary evolution operator, $\hat{U}(\phi)$, the  output state in the Schr\"odinger picture is $\left| \psi^{out} (\phi) \right\rangle  = \hat{U}(\phi) \left| \psi ^{in}  \right\rangle  $, where $\left| \psi^{in}  \right\rangle$ is the Heisenberg picture input state.  In Eq.~(\ref{MeasurementProbability}), the sum over $k$ is now a double sum over all non-negative values of the two integers $n_c$ and $n_d$.   For a generic Mach-Zehnder interferometer, with input ports ``a" and ``b", and output ports ``c" and ``d",  the projective operators are~\cite{Bahder2006}
\begin{equation}
\hat \Pi _\phi  \left( {n_c ,n_d } \right) = \frac{1}{{n_c !\,n_d !}}\,\left( {\hat c^\dag  } \right)^{n_c } \,( {\hat d^\dag  } )^{n_d } \,\,\left| 0 \right\rangle \,\left\langle 0 \right|\,\,\left( {\hat c} \right)^{n_c } \, ( {\hat d} )^{n_d } 
\label{POVM-MZ-2-Port}
\end{equation}
where the vacuum state $\left| 0 \right\rangle = \left| 0 \right\rangle_a \,\otimes \, \left| 0 \right\rangle_b  $.

As an example, consider the simplest case of input given by a mixed state represented by the density matrix
\begin{equation}
\hat \rho  = P_0 \,\left| 0 \right\rangle \left\langle 0 \right| + P_1 \,\left| {10} \right\rangle \left\langle {10} \right|
\label{1-photonDensityOperator}%
\end{equation}
where $P_1$ is the  probability for 1-photon input into port ``a" and vacuum input into port ``b", and  $P_0$ is the probability of vacuum input into both ports ``a" and ``b", and $P_0+ P_1=1$.  
In Eqs.~(\ref{DensityMatrix}) and (\ref{MeasurementProbability}), $P_S(\psi^{in})=P_o$ when $\left| \psi^{in} \right\rangle = \left| 0 \right\rangle $ and $P_S(\psi^{in})=P_1$ when $\left| \psi^{in} \right\rangle = \left| 10 \right\rangle $.
\begin{figure}[t]
\includegraphics[width=3.2in]{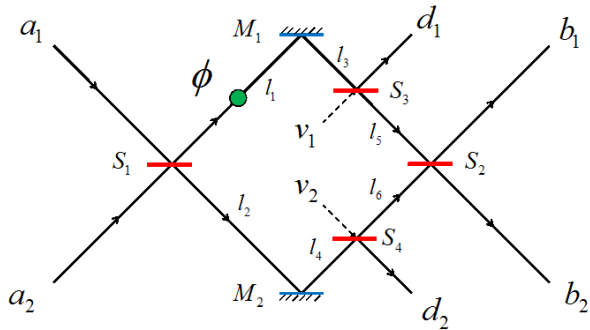}
\caption{\label{fig:LossyMachZehnder}(Color) Lossy Mach-Zehnder interferometer is shown, with input modes $a_1$, $a_2$, $v_1$, $v_2$, and output modes  $b_1$, $b_2$, $d_1$ and $d_2$.  
Here modes $v_1$ and $v_2$ have vacuum input and  the output modes $d_1$ and $d_2$ are loss channels in each arm.}  
\end{figure}

Equation~(\ref{MeasurementOps}) for the interferometer transfer matrix can then be written as
\begin{equation}
P(n_c ,n_d |\phi, \rho ) = {\rm{tr}}\left( {\hat \rho \,\hat \Pi _\phi  \left( {n_c ,n_d } \right)} \right)
\label{MeasurementProbabilityDensitymatrix2}
\end{equation}
where the input state is represented by the density matrix $\hat{\rho}$.
It seems that there can be only two possible measurement outcomes, $\xi = \{ n_c, n_d\} = \{ 1,0\}$ and $\xi = \{ n_c, n_d\} = \{ 0,1\}$.
However,  the probabilities for the two measurement outcomes do not sum to unity because, $P(10|\phi, \rho)+P(01|\phi, \rho)=P_1$. Therefore, there is a non-zero probability of an inconclusive measurement outcome associated with the probability $P_0$ of vacuum injected into both input ports ``a" and ``b".  I introduce an inconclusive measurement operator, $\hat \Pi _\phi  \left( i \right)$, so that the sum of the three operators is equal to the identity operator $\hat{I}$:
\begin{equation}
\hat \Pi _\phi  \left( {10} \right) + \hat \Pi _\phi  \left( {01} \right) + \hat \Pi _\phi  \left( i \right) = \hat{I}
\label{SumOfThreeMeasurementOps}%
\end{equation}
Using Eqs.(\ref{POVM-MZ-2-Port})--(\ref{SumOfThreeMeasurementOps}),  the probabilities given by Eq.~(\ref{MeasurementProbability}), $P(\xi | \phi )$, for  measurement  outcomes $\xi$ are given by
\begin{eqnarray}
P(10|\phi ) & = & P_1 \cos^2 \left( \frac{\phi}{2} \right) \label{MeasurementOutcomeProbabilityNonIdealStatepreparation_1} \\ 
P(01| \phi) & = & P_1 \sin ^2 \left(\frac{\phi}{2} \right) \label{MeasurementOutcomeProbabilityNonIdealStatepreparation_2} \\ 
P(i|\phi) & = & 1-P_1 
\label{MeasurementOutcomeProbabilityNonIdealStatepreparation_3}
\end{eqnarray}
where $P(i|\phi )$ is the probability for an inconclusive measurement outcome. 

Using Bayes' rule and Eq.~(\ref{MeasurementProbability}), we can write the conditional probability distributions for the phase, $p(\phi |\xi )$, given  measurement outcome, $\xi$, as
\begin{equation}
p(\phi |\xi ) = \frac{{P(\xi|\phi  )\,p(\phi )}}{{\int\limits_{ - \pi }^{ + \pi } {P(\xi|\phi '  )\,p(\phi ')\,d\phi '} }}
\label{PhaseProbabilityDistribution}
\end{equation}
where $p(\phi)$ is the prior probability distribution specifying our prior information about the phase $\phi$.
Assuming no prior information about the phase, $p(\phi)=1/(2 \pi)$, and using Bayes' rule in Eq.~(\ref{PhaseProbabilityDistribution}), the conditional  probability distributions for the phase for a given measurement outcome are given by
\begin{eqnarray}
 p(\phi |10) & = &\frac{1}{\pi} \cos^2  \left(\frac{\phi}{2} \right)   \label{MeasurementOutcomeProbabilityNonIdealStatepreparation-1}  \\ 
 p(\phi |01) & = & \frac{1}{\pi} \sin^2  \left(\frac{\phi}{2} \right)  \label{MeasurementOutcomeProbabilityNonIdealStatepreparation-2} \\ 
 p(\phi |i) & = & \frac{1}{{2\pi }}   \label{MeasurementOutcomeProbabilityNonIdealStatepreparation-3}
\end{eqnarray}
where $\xi=(n_c, n_d)$.
As we would expect, for an inconclusive measurement outcome the phase probability distribution, $p(\phi |i)$, is flat since an inconclusive measurement result cannot be used to distinguish different values of the phase.  Note that the probabilities for the three measurement outcomes in Eq.~(\ref{MeasurementOutcomeProbabilityNonIdealStatepreparation-1})--(\ref{MeasurementOutcomeProbabilityNonIdealStatepreparation-3}) sum to unity. When $P_0=0$, or equivalently $P_1=1$, state preparation is deterministic, and the probability for an inconclusive measurement outcome is zero.  In this case, the probabilities for measurement outcomes $(10)$ and $(01)$ reduce to the values for the case of a single-photon input state created with probability unity.  Note that this single photon input case  corresponds to the case of a classical interferometer fed by a laser in one input port.  Larger photon-number input states have measurement outcome probabilities  that differ from the probabilities for a classical interferometer.

For the input state in Eq.~(\ref{1-photonDensityOperator}), the classical Fisher information, defined in Eq.~(\ref{ClassicalFisherInformation}), is given by
\begin{equation}
F(\phi)= P_1 
\label{FisherNonDeterministicStatepreparation}%
\end{equation}
where I dropped the subscript ${cl}$ on the classical Fisher information, a convention that I follow in rest of this work. 
According to the Cramer-Rao bound in Eq.~(\ref{Cramer-RaoBound}), when the probability  of creating a single photon approaches zero,  $P_1 \rightarrow 0$,  the variance $(\delta\phi)^2$ becomes arbitrarily large, because  the probability for inconclusive measurement outcomes approaches unity.

Assuming no prior information about the phase, therefore taking $p(\phi)=1/( 2\pi)$, the  fidelity  (Shannon mutual information) of the system defined in Eq.~(\ref{ShannonMutualInformation}) is given by
\begin{equation}
H(M) =  P_1 \,\left( {\frac{1}{{\ln 2 }} - 1} \right)
\label{Fidelity-Non-IdealDetector}
\end{equation}
Similar to the Fisher information, the fidelity $H(M)$ also approaches zero when the probability $P_1$ of having one photon in the input of each shot approaches zero.  The fidelity  is the amount of information (in bits) that is gained on average about the phase from a single use of the interferometer, averaged over all possible phase values $\phi$.

\subsection{\label{LossyInterferometer}Lossy Mach-Zehnder Interferometer}
Next, I consider an interferometer with absorption losses---so state evolution is non-unitary.   I assume that state preparation is deterministic (ideal) and that state detection is perfect (no errors).  Equation~(\ref{MeasurementProbability}) is general enough to describe  processes other than losses in the interferometer, such as photons scattering into the interferometer from the environment, in which  case there are more photons leaving the output ports than entering the input ports. However, in what follows, I restrict myself to simple absorption in the interferometer. I model losses in each arm of a Mach-Zehnder interferometer by inserting two beam splitters, $S_3$ and $S_4$, one in each path, see Fig.~\ref{fig:LossyMachZehnder}.  While a lossless Mach-Zehnder interferometer has two input and two output ports, a general lossy Mach-Zehnder interferometer can be represented by four input and four output ports, see Fig.~\ref{fig:LossyMachZehnder}. I label the input modes as $a_1$, $a_2$, $v_1$, and $v_2$, where $v_1$, and $v_2$ have vacuum input and I label the output modes as $b_1$, $b_2$, $d_1$, and $d_2$, where $d_1$, and $d_2$ are the modes where probability amplitude is ``dissipated". I take the phase shifts at the two mirrors, $M_1$ and $M_2$ to be equal to $\pi$.  Furthermore, I assume that the interferometer is balanced, so that path lengths satisfy, 
\begin{eqnarray}
L & =  & l_1+l_3+l_5=l_2+l_4+l_6 \nonumber \\  
l & =  &l_1+l_3=l_2+l_4   \label{PathLengths}
\end{eqnarray}
see Fig.~\ref{fig:LossyMachZehnder}.  A calculation gives the input and output modes related by the 4$\times$4 unitary scattering matrix $S_{ij}(\phi)$
\begin{equation}
\left( {\begin{array}{*{20}c}
   {b_1 }  \\
   {b_2 }  \\
   {d_1 }  \\
   {d_2 }  \\
\end{array}} \right) = \left[ {\begin{array}{*{20}c}
   {} & {} & {}  \\
   {} & {S_{ij} \left( \phi  \right)} & {}  \\
   {} & {} & {}  \\
\end{array}} \right] \cdot \left( {\begin{array}{*{20}c}
   {a_1 }  \\
   {a_2 }  \\
   {v_1 }  \\
   {v_1 }  \\
\end{array}} \right)
\label{InOutRelationsLossyMZ}
\end{equation}
where the phase-dependent scattering matrix  is given by
\begin{widetext}
\begin{equation}
S_{ij} (\phi ) = \left[ {\begin{array}{*{20}c}
   {\frac{i}{2}e^{i\frac{{L\omega }}{c}} \left( {\sqrt {1 - r_y^2 }  - e^{i\phi } \sqrt {1 - r_x^2 } } \right)} & { - \frac{1}{2}e^{i\frac{{L\omega }}{c}} \left( {\sqrt {1 - r_y^2 }  + e^{i\phi } \sqrt {1 - r_x^2 } } \right)} & {\frac{i}{{\sqrt 2 }}r_x \,e^{i\left( {L - l} \right)\frac{\omega }{c}} } & {\frac{1}{{\sqrt 2 }}r_y \,e^{i\left( {L - l} \right)\frac{\omega }{c}} }  \\
   { - \frac{1}{2}e^{i\frac{{L\omega }}{c}} \left( {\sqrt {1 - r_y^2 }  + e^{i\phi } \sqrt {1 - r_x^2 } } \right)} & { - \frac{i}{2}e^{i\frac{{L\omega }}{c}} \left( {\sqrt {1 - r_y^2 }  - e^{i\phi } \sqrt {1 - r_x^2 } } \right)} & {\frac{1}{{\sqrt 2 }}r_x \,e^{i\left( {L - l} \right)\frac{\omega }{c}} } & {\frac{i}{{\sqrt 2 }}r_y \,e^{i\left( {L - l} \right)\frac{\omega }{c}} }  \\
   { - \frac{i}{{\sqrt 2 }}r_x \,e^{i\left( {\phi  + \frac{{l\omega }}{c}} \right)} } & { - \frac{1}{{\sqrt 2 }}r_x \,e^{i\left( {\phi  + \frac{{l\omega }}{c}} \right)} } & { - i\sqrt {1 - r_x^2 } } & 0  \\
   { - \frac{1}{{\sqrt 2 }}r_y \,e^{i\frac{{l\omega }}{c}} } & { - \frac{i}{{\sqrt 2 }}r_y \,e^{i\frac{{l\omega }}{c}} } & 0 & { - i\sqrt {1 - r_y^2 } }  \\
\end{array}} \right]
\label{LossyMZScatteringMatrix}%
\end{equation}
\end{widetext}
It is easy to check that the scattering matrix is unitary, $S^\dag  \,S = I$ where $I$ is the 4$\times$4 unit matrix. The parameters,  $r_x$ and $r_y$, are the reflection amplitudes for beams splitters $S_3$ and $S_4$, respectively, and they represent the strength of the loss or dissipation, see Fig.~\ref{fig:LossyMachZehnder}.  When the system is considered in terms of two input modes, $a_1$ and $a_2$, and two output modes, $b_1$ and $b_2$, the evolution of the input state in not unitary.  

The 4$\times$4 unitary scattering matrix $S_{ij} (\phi )$ has some simple properties.   The case when $r_x=r_y=0$ corresponds to no dissipation. In this case, the 4$\times$4 S-matrix reduces to two diagonal 2$\times$2 blocks.  The upper left 2$\times$2 block couple modes $a_1$ and $a_2$ to modes $b_1$ and $b_2$, and this 2$\times$2 block  (up to a phase)  is given by Eq.~(\ref{Smatrix2x2}), which is the scattering matrix for the Mach-Zehnder interferometer with no losses. For this case of no loss, in Eq.~(\ref{LossyMZScatteringMatrix}) the lower right 2$\times$2 block couples the dissipative modes, $d_1$, and $d_2$, to the vacuum modes, $v_1$, and $v_2$. 

The case $r_x=r_y=1$ corresponds to maximum dissipation, and the 4$\times$4  S-matrix again decouples, into two off-diagonal 2$\times$2 blocks.  The upper right 2$\times$2 block couples the two vacuum modes, $v_1$ and $v_2$, to the two output modes,  $b_1$ and $b_2$.  The lower left 2$\times$2 block  of this S-matrix couples the loss  modes, $d_1$, and $d_2$, to the input modes $a_1$ and $a_2$.  For this case of maximum dissipation, the input  modes, $a_1$ and $a_2$, are  decoupled from the  output modes, $b_1$ and $b_2$.

The probabilities for various measurement outcomes $\xi=(n , m)$ are given by the analog of Eq.~(\ref{MeasurementProbabilityDensitymatrix}):
\begin{equation}
P(n,m|\phi ,\rho ) = {\rm{tr}}\left( {\hat \rho _o \,\hat \Pi _\phi ^{} (n,m)} \right) 
\label{MeasurementOutcomes2}
\end{equation}
where $n$ and $m$ are the number of photons leaving ports $b_1$ and $b_2$, respectively.  The trace is over the complete space of four direct product Fock basis states, 
${ \left| n_1 \right\rangle _{a_1 }  \otimes \left| n_2 \right\rangle _{a_2 }  \otimes \left| n_3 \right\rangle _{v_1 }  \otimes \left| n_4 \right\rangle _{v_2 } }$.  
The input state  density matrix, $\hat \rho _o$, is defined in terms of a sum of products of creation operators $\hat{a}_1^\dag $, $\hat{a}_2^\dag $, $\hat{v}_1^\dag $ and $\hat{v}_2^\dag $ acting on the vacuum $ \left| 0 \right\rangle  = \left| 0 \right\rangle _{a_1 }  \otimes \left| 0 \right\rangle _{a_2 }  \otimes \left| 0 \right\rangle _{v_1 }  \otimes \left| 0 \right\rangle _{v_2 } $   and has the form
\begin{equation}
\hat \rho _0  = \sum\limits_{n,m} {c_{nm} \,\left| {nm00} \right\rangle } \;\left\langle {nm00} \right|
\label{InputDensitymatrix}
\end{equation} 
where I use the short-hand notation 
${ \left| {nm00} \right\rangle  \equiv \left| n \right\rangle _{a_1 }  \otimes \left| m \right\rangle _{a_2 }  \otimes \left| 0 \right\rangle _{v_1 }  \otimes \left| 0 \right\rangle _{v_2 } }$.
I am using the Heisenberg picture, so the input density matrix $\hat \rho _o$ is independent of time (phase), while the operators $\hat \Pi _\phi  \left( {n,m} \right)$ evolve in time (phase) and so they depend on $\phi$.  The projective measurement operators are given by 
\begin{widetext}
\begin{equation}
\hat \Pi _\phi  \left( {n,m} \right) = \frac{1}{{n!\,m!}}\;\sum\limits_{k,l = 0}^\infty  {\frac{1}{{k!\,l!}}\left( {\hat b_1 ^\dag  } \right)^n \left( {\hat b_2 ^\dag  } \right)^m \left( {\hat d_1 ^\dag  } \right)^k \left( {\hat d_2 ^\dag  } \right)^l \left| 0 \right\rangle \left\langle 0 \right|\left( {\hat b_1 } \right)^n \left( {\hat b_2 } \right)^m \left( {\hat d_1 } \right)^k \left( {\hat d_2 } \right)^l } 
\label{ProbabilityMN}
\end{equation}
\end{widetext}
where $\hat b_1 ^\dag$,  $\hat b_2 ^\dag$, $\hat d_1 ^\dag$, and $\hat d_2 ^\dag$ are creation operators for the output modes $b_1$, $b_2$, $d_1$, and $d_2$, respectively.
The sums over $k$ and $l$ take into account the probabilities for losing photons into ports $d_1$ and $d_2$.  
I use the short-hand notation for the input modes 
$\{ \alpha_i \} = \{ \hat{a}_1 ,\hat{a}_2,\hat{v}_1,\hat{v}_2 \}$ and for the output modes $\{ \beta_i \} = \{ \hat{b}_1 ,\hat{b}_2,\hat{d}_1,\hat{d}_2 \}$, see Eq.~(\ref{InputOutputDef}).

Note that, even though dissipation is being modeled, it is easy to check that there is global photon number conservation, 
\begin{equation}
\sum\limits_{i = 0}^4 {\hat{\alpha}_i^\dag  \hat{\alpha}_i  = } \sum\limits_{i = 0}^4 {\hat{\beta}_i^\dag  \hat{\beta}_i } 
\label{PhotonConservation}
\end{equation}
since the S-matrix in Eq.~(\ref{LossyMZScatteringMatrix}) is unitary. 
The measurement outcomes, $\xi=(n,m)$, are specified by two integers, which label the number of photons that are output at ports $b_1$ and $b_2$, see Fig.~\ref{fig:LossyMachZehnder}. 
So the probability of output state, $\psi _j^{out}$, as given in Eq.(\ref{MeasurementOutcomes2}), is expressed by the two integers $n$ and $m$, see also Eq.~(\ref{MeasurementOps}).  Note that in general the number of photons $n+m$  in the output state, $\psi _j^{out}$, is not equal to the number of photons in the input state, $\psi_{in}$, because
\begin{equation}
\hat{a}_1^\dag  \hat{a}_1  + \hat{a}_2^\dag  \hat{a}_2  \ne \hat{b}_1^\dag  \hat{b}_1  + \hat{b}_2^\dag  \hat{b}_2 
\label{noPhotonConservation}
\end{equation} 
However, the sum of probabilities for all possible measurement outcomes is unity:
\begin{equation}
\sum\limits_{m,n = 0}^\infty  {P(n,m|\phi ,\rho )}  = 1
\label{ProbabilitySumUnity}
\end{equation}
where $P(n,m|\phi ,\rho )$  is given by Eq.~(\ref{MeasurementOutcomes2}), and is a result of 
\begin{equation}
\sum\limits_{n,m} {\hat \Pi _\phi  \left( {n,m} \right)}  = \hat{I}
\label{OpertorsSumToUnity}
\end{equation}

For the case of a pure input state, $\left| {\psi _{in} } \right\rangle $, it is convenient to define the operators
\begin{equation}
\hat N\left( {n,m,k,l} \right) = \frac{1}{{n!\,m!\,k!\,l!}}\left( {\hat b_1 } \right)^n \left( {\hat b_2 } \right)^m \left( {\hat d_1 } \right)^k \left( {\hat d_2 } \right)^l 
\label{Nops}
\end{equation}
and the probabilities in Eq.~(\ref{MeasurementOutcomes2}) are then given by
\begin{equation}
P(n,m|\phi ,\psi _{in} ) = \sum\limits_{k,l = 1}^\infty  {\left| {\left\langle 0 \right|\hat N\left( {n,m,k,l} \right)\left| {\psi _{in} } \right\rangle } \right|} ^2 
\label{MeasProbs}
\end{equation}

Equation~(\ref{MeasurementOutcomes2}), or Eq.~(\ref{MeasProbs}) for the case of pure states, defines a unitary mapping,  $|nm00 \rangle \rightarrow |n^\prime m^\prime k l \rangle$, between  interferometer input states,   $  |nm00 \rangle$, and output states, $|n^\prime m^\prime k l \rangle $, because the photon number is conserved: 
${ n+m =n^\prime + m^\prime + k + l }$, see Eq.~(\ref{PhotonConservation}).   

If we restrict our attention to measurement outcomes projected onto the Hilbert subspace with  basis  
$ { \left| n^\prime \right\rangle _{b_1 }  \otimes \left| m^\prime \right\rangle _{b_2 } }$, 
Eq.~(\ref{MeasurementOutcomes2}) or Eq.~(\ref{MeasProbs}) defines the non-unitary mapping ${\cal E}$:
\begin{equation}
{\cal E} \left[ \left| n \right\rangle _{a_1 }  \otimes \left| m \right\rangle _{a_2 }  \right] \rightarrow \left| n^\prime \right\rangle _{b_1 }  \otimes \left| m^\prime \right\rangle _{b_2 }  
\label{Non-UnitaryMapping}
\end{equation}
since photon number is not conserved: $n+m \ne n^\prime + m^\prime $, see Eq.~(\ref{noPhotonConservation}), which represents losses in the Mach-Zehnder interferometer. Here the Fock states ${| n^\prime \rangle _{b_i }}$, for $i=1,2$, are created from the vacuum state, $| 0 \rangle _{b_i }$, by application of creation operators $ \hat{b}_i^\dag $ in the usual way.  The mapping ${\cal E}$ depends on two parameters, $r_x$ and $r_y$, which specify the strength of the dissipation or losses in each arm of the interferometer.  In the limit of no dissipation, when $r_x=0$ and $r_y=0$, the mapping ${\cal E}$ becomes a unitary transformation and $n+m=n^\prime + m^\prime$.

In what follows, I use the short-hand notation 
$ | n m \rangle$ for the input state $| n m 0 0 \rangle \equiv   | n \rangle_{a_1}  \otimes  | m \rangle_{a_2}  \otimes  | 0 \rangle_{v_1} \otimes  | 0 \rangle_{v_2}  $.

In the next two subsections, \ref{FockStateInput} and \ref{N00NStateInputInput}, I discuss the Fisher information and the fidelity (Shannon mutual information) for specific cases of few-photon  Fock state and N00N state input into the lossy Mach-Zehnder (MZ) interferometer.

\subsection{\label{FockStateInput}Fock State Input into Lossy MZ Interferometer}

Consider the $N$-photon Fock state 
\begin{equation}
\left| {\psi _N } \right\rangle  = \frac{1}{{\sqrt {N!} }}\left( {\hat a_1^\dag  } \right)^N \,\left| 0 \right\rangle  = \left| {N000} \right\rangle \equiv   \left| {N0} \right\rangle 
\label{N-PhotonInputState}
\end{equation}
input into a lossy Mach-Zehnder interferometer given by the scattering matrix in Eq.~(\ref{LossyMZScatteringMatrix}).  The  probabilities for measurement outcomes $\xi=(n,m)$ are given by:
\begin{eqnarray}
 & & P(n , m | \phi ,\psi_N ) =   
 \frac{N!}{n!m!} \sum\limits_{k = 0}^N  \sum\limits_{l = 0}^N \frac{1}{k!l!}  \times  \nonumber \\
 &  & \left| S_{11}^n S_{21}^m S_{31}^k S_{41}^l  \right|^2 \,\delta_{n + m + k + l,N}   
\label{MeasurementProbsForFOCK} 
\end{eqnarray}
where $S_{ij}$ are the matrix elements of Eq.~(\ref{LossyMZScatteringMatrix}) and $\delta_{m, n}$ is the Kronecker delta function.
 
A direct calculation of the classical Fisher information for the $N$-photon Fock state input, $F_N(\phi)$, gives  
\begin{equation}
F_N(\phi) = N F_1(\phi)
\label{N-photonFisherInfo}
\end{equation} 
where $F_1(\phi)$ is the classical Fisher information for one-photon input, given in Eq.~(\ref{Fisher-1-Photon}).  This shows that for the lossy MZ interferometer with  $N$-photon  Fock state input, the standard deviation $(\delta\phi)$ scales as $1/ \sqrt{N}$.  From another point of view, since the Fisher information is additive for independent events, the $N$-photon Fock state acts like $N$ independent 1-photon states.   When $r_x=r_y$, then $F_1(\phi)=1$, and the $N$-photon Fisher information becomes $F_N(\phi)=N$. This means that dissipation in the (non-unitary) interferometer has the effect of introducing a phase dependence into the Fisher information, see the discussion below.

\subsubsection{\label{1-PhotonFockInput}1-Photon Fock State Input into Lossy MZ Interferometer}

As the simplest example of the effect of dissipation, I consider the 1-photon Fock state input into the lossy Mach-Zehnder interferometer with scattering matrix given by Eq.~(\ref{LossyMZScatteringMatrix}) 
\begin{equation}
\left| \psi_{in} \right\rangle=  \hat{a}_1^{\dag} \left| 0 \right\rangle =   \left| 1000 \right\rangle  \equiv  \left| 10 \right\rangle
\label{1-photonFockState}
\end{equation}
where I use the short-hand notation $\left| 10 \right\rangle$ for the input state $\left| 1000 \right\rangle$.
The probabilities for measurement outcomes are given by $P( n , m |\phi ,\psi _{in} )$, where $n , m$ specify the photon numbers output in port $b_1$ and $b_2$, respectively, see Fig.~\ref{fig:LossyMachZehnder}.    The probabilities  $P(n, m |\phi ,\psi _{in} )$ for the three measurement outcomes are:
\begin{equation}
\small
\begin{array}{lllll}
 P(10|\phi , 10 ) &  =  & \frac{1}{4}\left( {2 - r_x^2  - r_y^2  - 2\sqrt {\left( {1 - r_x^2 } \right)\left( {1 - r_y^2 } \right)} \,\cos \phi } \right) \\ 
 P(01|\phi , 10  ) &  =  & \frac{1}{4}\left( {2 - r_x^2  - r_y^2  + 2\sqrt {\left( {1 - r_x^2 } \right)\left( {1 - r_y^2 } \right)} \,\cos \phi } \right) \\ 
 P(00|\phi , 10  ) &  =  & \frac{1}{2}\left( {r_x^2  + r_y^2 } \right) \\ 
 \end{array}
\label{1-PhotonOutcome}
\end{equation}
The probability $P(0 0|\phi , 1 0 )$  is associated with an inconclusive measurement outcome, since for this case zero photons leave the output ports, i.e., the photon that entered in port ``a"  was absorbed in the interferometer, or more precisely the photon was output in either port $d_1$ or $d_2$.

\begin{figure}[t]   
\includegraphics[width=3.4in]{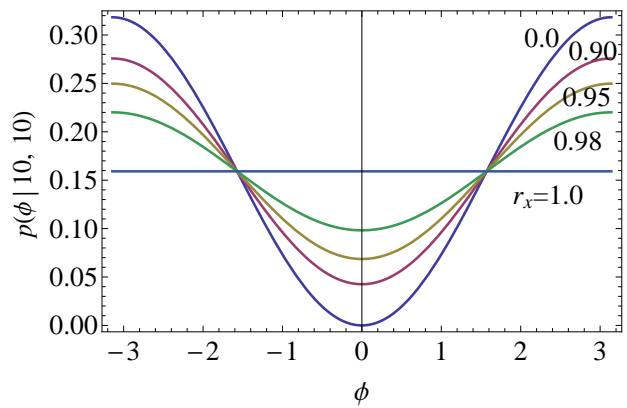}
\caption{\label{fig:MZLossy-ContrastLoss} (Color) For the 1-photon input state $\left| 10 \right\rangle$, given by Eq.~(\ref{1-photonFockState}),  the probability distribution for the phase, $p(\phi|10, 10)$, in Eq.~(\ref{1-PhotonPhaseDistributions}) is plotted for absorption $r_y=0$ and $r_x =$0.0, 0.90, 0.95, 0.98, and 1.0. As  $r_x \rightarrow$1.0, the probability distribution  becomes flat and  does not distinguish between different phase values. }
\end{figure}

From Bayes' rule in Eq.~(\ref{PhaseProbabilityDistribution}), the phase probability distributions, $ p(\phi| m \, n, \psi_{in}) $, for input state 
$\left| \psi_{in}  \right\rangle $  given in Eq.~(\ref{1-photonFockState}), are given by
\begin{equation}
\begin{array}{lllll}
 p(\phi |10, 10) &  =  & \frac{1}{{2\pi }}\frac{{2 - r_x^2  - r_y^2  - 2\sqrt {\left( {1 - r_x^2 } \right)\left( {1 - r_y^2 } \right)} \,\cos \phi }}{{2 - r_x^2  - r_y^2 }} \\ 
 p(\phi |01, 10) &  =  & \frac{1}{{2\pi }}\frac{{2 - r_x^2  - r_y^2  + 2\sqrt {\left( {1 - r_x^2 } \right)\left( {1 - r_y^2 } \right)} \,\cos \phi }}{{2 - r_x^2  - r_y^2 }} \\ 
 p(\phi |00, 10) &  =  & \frac{1}{{2\pi }} \\ 
 \end{array}
\label{1-PhotonPhaseDistributions}
\end{equation}
When $r_x \ne r_y$, there is a loss of contrast in the phase probability distributions $ p(\phi| m \, n, \psi_{in}) $, see Fig.~\ref{fig:MZLossy-ContrastLoss}.
When the absorption probabilities are the same in both arms, in the limit $r_x = r_y$,  the phase probability distributions in Eq.~(\ref{1-PhotonPhaseDistributions}) reduce to the case of a single photon input without losses in the interferometer, which are given in 
Eq.~(\ref{MeasurementOutcomeProbabilityNonIdealStatepreparation-1})-(\ref{MeasurementOutcomeProbabilityNonIdealStatepreparation-3}), with trivial phase change given by the replacements $\sin \rightarrow \cos$. The phase probability density, $p(\phi | 00, 10)$, is associated with the inconclusive outcome, $p(\phi|i)$.  It is a remarkable feature that for equal loss in both arms, $r_x=r_y$, the phase probability densities, 
$p(\phi| m \, n, \psi_{in}) $, in Eq.(\ref{1-PhotonPhaseDistributions}) do not depend on the size of the loss, $r_x$.  However, there is loss of information with increasing absorption, $r_x$ and $r_y$, which is reflected in the information measures, see below.    

The effect of equal dissipation in both arms of the interferometer is the same as the effect of non-deterministic state preparation,  specified by  input state characterized by a density matrix in Eq.~(\ref{1-photonDensityOperator}).  However, when the dissipation in both arms is not equal, say for $r_y=0$  and $r_x$ is finite, the phase probability distributions show a loss of contrast, see Fig.~\ref{fig:MZLossy-ContrastLoss}.  This feature may be useful in applications to null-type measurements.
\begin{figure}[t]   
\includegraphics[width=3.2in]{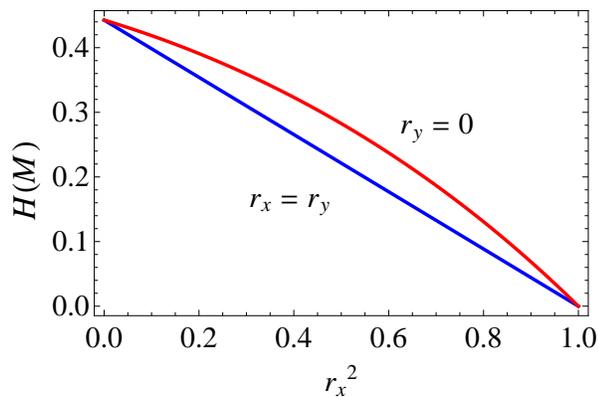}
\caption{\label{fig:H-1-photonLinerPlot} (Color) For 1-photon Fock  input state, $\left| 10 \right\rangle$, given by Eq.~(\ref{1-photonFockState}), the fidelity (Shannon mutual information) $H(M)$ seems linear in $r_x^2$ for $r_x=r_y$, however, this is not true for general values of $r_x$ and $r_y$, see also Fig.~\ref{fig:H-1-photonLinerPlot3D}.}
\end{figure}
\begin{figure}[t]   
\includegraphics[width=3.6in]{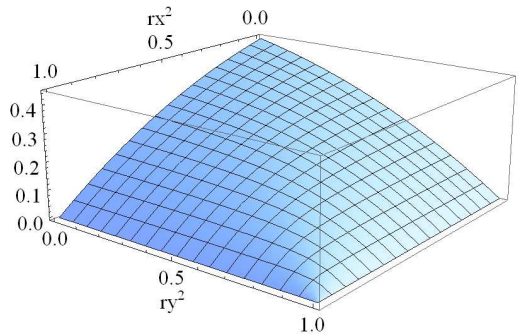}
\caption{\label{fig:H-1-photonLinerPlot3D} (Color) For 1-photon Fock  input state  $\left| 1000 \right\rangle = \left| 10 \right\rangle$, given by Eq.~(\ref{1-photonFockState}), the fidelity (Shannon mutual information) $H(M)$ is plotted as a function of loss parameters $r_x^2$ and $r_y^2$.}
\end{figure}
\begin{figure*}   
\includegraphics[width=6.5in]{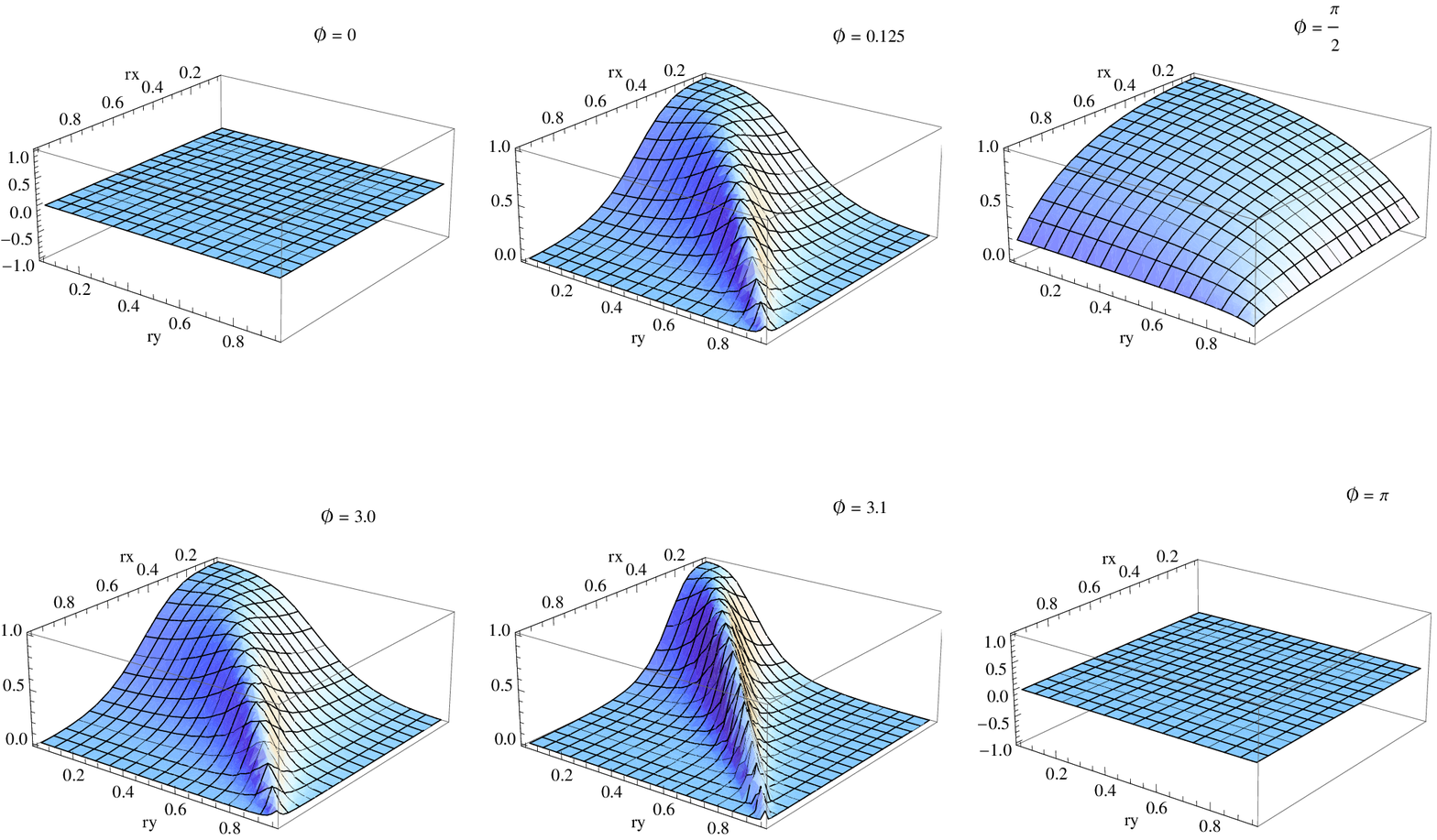}
\caption{\label{fig:Fisher-1-Photon-Lossy-Interferomter} (Color) The Fisher information $F(\phi)$ for 1-photon input state  $\left| 1000 \right\rangle$ (short-hand notation   $\left| 10 \right\rangle$), is plotted as a function of $r_x$ and $r_y$, for different values of $\phi= 0, 0.125, \pi/2, 3.0, 3.1, \pi$, left to right in top row and bottom row.}
\end{figure*}

In the discussion that follows, I assume no prior information on the phase, so I take $p(\phi)=1/(2 \pi)$.  When there is no loss in the interferometer, $r_x=r_y=0$,  the Shannon mutual information (fidelity) as defined in Eq.~(\ref{ShannonMutualInformation}) is a constant:  
\begin{equation}
H( M ) = \frac{1}{{\ln 2}} - 1
\label{1-PhotonFidelity-rx=ry=0}
\end{equation}
When the losses in both arms are equal, $r_x=r_y$, we have the exact result 
\begin{equation}
H( M ) = \left( {\frac{1}{{\ln 2}} - 1} \right)\left( {1 - r_x^2 } \right)
\label{1-PhotonFidelity-rx=ry}
\end{equation}
For general values of $r_x$ and $r_y$, the expression for $H(M)$ is large and complicated, but for small ${r_x \ll 1}$ and ${r_y \ll 1}$, I can expand it in a power series,  
\begin{equation}
\small
H( M)  = \left( {\frac{1}{{\ln 2}} - 1} \right)\left[ {1 - \frac{1}{2}\left( {r_x^2  + r_y^2 } \right)} \right]  + O(r_x^4 ) + O(r_y^4 )
\label{1-PhotonFidelitySmallRxRy}
\end{equation}
where I dropped fourth order terms in $r_x$ and $r_y$.
When $r_x=r_y$, from Eq.~(\ref{1-PhotonFidelitySmallRxRy}), we may expect the fidelity (Shannon mutual information) $H(M)$ to be quadratic in $r_x$,  however, this is not true for general values $r_x$ and $r_y$, see Fig.~\ref{fig:H-1-photonLinerPlot}, which shows $H( M)$ vs. $r_x^2$ for the case $r_x=r_y$ and for $r_y=0$.  

In Figure~\ref{fig:H-1-photonLinerPlot3D} the fidelity (Shannon mutual information) $H(M)$ is plotted as a function of the dissipation parameters,  $r_x$ and $r_y$.   When the dissipation in either arm is a maximum, $r_x=1$ or $r_y=1$,  the fidelity $H(M)= 0$, indicating that we obtain zero information from each photon.

For the 1-photon Fock state  (given in Eq.~(\ref{1-photonFockState})) input into the lossy Mach-Zehnder interferometer given in Eq.~(\ref{LossyMZScatteringMatrix}), the classical Fisher information (defined by Eq.~(\ref{ClassicalFisherInformation})) is given by:
\begin{equation}
F_1(\phi)=\frac{2 \left(1-r_x^2\right) \left(1-r_y^2\right)
   \left(2 -r_x^2-r_y^2 \right) \sin ^2(\phi
   )}{\left(2 -r_x^2-r_y^2 \right)^2-4 \left(1-r_x^2\right)
   \left(1-r_y^2\right) \cos ^2(\phi )}
\label{Fisher-1-Photon}
\end{equation}
see plots in Fig.~\ref{fig:Fisher-1-Photon-Lossy-Interferomter}.  From these plots, it is clear that for a lossy interferometer, the  Fisher information depends strongly on the true value of $\phi$.   Through the  Cramer-Rao bound in Eq.~(\ref{Cramer-RaoBound}), this translates to a dependence of the variance $(\delta\phi)^2$ on the true value of $\phi$.

Therefore, Fisher information $F_1(\phi)$ for 1-photon Fock state input into a lossy interferometer is qualitatively different than for an ideal interferometer without loss, see Eq.~(\ref{Fisher-1-Photon}) for the case $r_x = r_y = 0$.  As a consequence of Eq.(\ref{N-photonFisherInfo}), the Fisher information for $N$-photon Fock state input into a lossy interferometer is qualitatively different than for a lossless interferometer, where it is a constant given by $F_N(\phi)=N$.     Specifically, for a lossy interferometer the Fisher information depends on the value of the true phase, see plots in Fig.~\ref{fig:Fisher-1-Photon-Lossy-Interferomter}.   For values of the phase given by $\phi=0$ and $\phi= \pm \pi$, the Fisher information vanishes for Fock state input, independent of the value of the dissipation parameters, $r_x$ and $r_y$, see comment~\footnote{This statement applies to a balanced interferometer, whose path lengths satisfy Eq.~(\ref{PathLengths}).}.  According to the Cramer-Rao bound,  the variance $(\delta\phi)^2 $, is large for values of true phase near $\phi=0$ and $\phi= \pm \pi$.   However, when there is no dissipation, $r_x=r_y=0$, the Fisher information is independent of $\phi$ and so is the bound on the variance $(\delta\phi)^2 $.  The presence of dissipation in the interferometer introduces a dependence of the variance, $(\delta\phi)^2 $, on true phase $\phi$.  The exception to this is when $r_x = r_y$, where $F_1(\phi)$ reduces to $F_1(\phi) = 1 - r_x^2$, and is independent of $\phi$.
%
%
%

\subsubsection{\label{2-PhotonFockInput}2-Photon Fock State Input into Lossy MZ Interferometer}
Consider now the 2-photon Fock state input into the lossy Mach-Zehnder interferometer with S-matrix given by  Eq.~(\ref{LossyMZScatteringMatrix}): 
\begin{equation}
\left| \psi_{in} \right\rangle   = \frac{1}{{\sqrt 2 }}\left( {a_1^\dag  } \right)^2  \left| 0 \right\rangle = \left| 2 0 00\right\rangle \equiv \left| 2 0 \right\rangle 
\label{2-photonInputState}
\end{equation} 
where I use the short-hand notation $\left| 2 0 \right\rangle$ for the state $\left| 2000\right\rangle$. The probabilities $P(m , n | \phi, \psi_{in})$ for the six measurement outcomes are given by
\begin{widetext}
\small
\begin{eqnarray}
P(20|\phi,20) & = & \frac{1}{{16}}\left[ {6 - 6r_x^2  - 6r_y^2  + 4\sqrt {\left( {r_x^2  - 1} \right)\left( {r_y^2  - 1} \right)} \left( {r_x^2  + r_y^2  - 2} \right)\cos (\phi ) + 2\left( {r_x^2  - 1} \right)\left( {r_y^2  - 1} \right)\cos (2\phi ) + 4r_x^2 r_y^2  + r_x^4  + r_y^4 } \right]   \nonumber \\
P(02|\phi,20 ) & = & \frac{1}{{16}}\left[ {6 - 6r_x^2  - 6r_y^2  - 4\sqrt {\left( {r_x^2  - 1} \right)\left( {r_y^2  - 1} \right)} \left( {r_x^2  + r_y^2  - 2} \right)\cos (\phi ) + 2\left( {r_x^2  - 1} \right)\left( {r_y^2  - 1} \right)\cos (2\phi ) + 4r_x^2 r_y^2  + r_x^4  + r_y^4 } \right]  \nonumber \\
P(11|\phi,20 )& = & \frac{1}{8}\left[ {2 - 2r_x^2  - 2r_y^2  + r_x^4  + r_y^4  - 2\left( {1 - r_x^2 } \right)\left( {1 - r_y^2 } \right)\cos \left( {2\phi } \right)} \right]  \nonumber  \\
P(10|\phi,20 )  & = &   \frac{1}{4}\left( {r_x^2  + r_y^2 } \right)\left[ {2 - r_x^2  - r_y^2  - 2\sqrt {\left( {1 - r_x^2 } \right)\left( {1 - r_y^2 } \right)} \,\cos \left( \phi  \right)} \right] \nonumber  \\
P(01|\phi,20 )  & = &  \frac{1}{4}\left( {r_x^2  + r_y^2 } \right)\left[ {2 - r_x^2  - r_y^2  + 2\sqrt {\left( {1 - r_x^2 } \right)\left( {1 - r_y^2 } \right)} \,\cos \left( \phi  \right)} \right]  \nonumber  \\
P(00|\phi,20) & = &   \frac{1}{4} \left( r_x^2 + r_y^2  \right)^2   \nonumber \\
\label{2-PhotonMeasurementProb}
\end{eqnarray}
\end{widetext}
The sum of the probabilities for the six possible measurement outcomes in Eq.~(\ref{2-PhotonMeasurementProb}) is unity. Since a two-photon Fock state has been used as input, the probability that no photon was absorbed is equal to the the sum  $P(20|\phi )+P(02|\phi )+P(11|\phi ) = \frac{1}{4}\left( {2 - r_x^2  - r_y^2 } \right)^2 $, while the probability that exactly one photon was absorbed is  $P(10|\phi )+P(01|\phi )=\frac{1}{2}\left( {2r_x^2  + 2r_y^2  - 2r_x^2 r_y^2  - r_x^4  - r_y^4 } \right)$.

The conditional probability distributions for the phase, $p(\phi | m \, n,  \psi_{in} )$ , with input state $ \left|  \psi_{in} \right \rangle $ in Eq.~(\ref{2-photonInputState}) and  measurement outcome $\xi = (m , n)$ are given by
\begin{widetext}
\begin{eqnarray}
p(\phi |20,20)  & = &  \frac{1}{{2\pi }}\left[ {1 +  \frac{{4\sqrt {\left( {1 - r_x^2 } \right)\left( {1 - r_y^2 } \right)} \left( {r_x^2  + r_y^2  - 2} \right)\cos (\phi ) + 2\left( {r_x^2  - 1} \right)\left( {r_y^2  - 1} \right)\cos (2\phi )}}{{6 - 6(r_x^2  + r_y^2 ) + r_x^4  + r_y^4  + 4r_x^2 r_y^2 }}} \right] \nonumber  \\
p(\phi |02,20) & = &  \frac{1}{{2\pi }}\left[ {1 - \frac{{4\sqrt {\left( {1 - r_x^2 } \right)\left( {1 - r_y^2 } \right)} \left( {r_x^2  + r_y^2  - 2} \right)\cos (\phi ) + 2\left( {r_x^2  - 1} \right)\left( {r_y^2  - 1} \right)\cos (2\phi )}}{{6 - 6(r_x^2  + r_y^2 ) + r_x^4  + r_y^4  + 4r_x^2 r_y^2 }}} \right]  \nonumber  \\
p(\phi |11,20) & = &  \frac{1}{{2\pi }}\left[ {1 - \frac{{2\left( {1 - r_x^2 } \right)\left( {1 - r_y^2 } \right)\cos (2\phi )}}{{2 - 2(r_x^2  + r_y^2 ) + r_x^4  + r_y^4 }}} \right] \nonumber  \\
p(\phi |10,20)  & = & \frac{1}{{2\pi }}\left[ {1 - \frac{{2\sqrt {\left( {1 - r_x^2 } \right)\left( {1 - r_y^2 } \right)} \cos (\phi )}}{{2 - r_x^2  - r_y^2 }}} \right] \nonumber  \\
p(\phi |01,20) & = &  \frac{1}{{2\pi }}\left[ {1 + \frac{{2\sqrt {\left( {1 - r_x^2 } \right)\left( {1 - r_y^2 } \right)} \cos (\phi )}}{{2 - r_x^2  - r_y^2 }}} \right]  \nonumber  \\
p(\phi |00,20) & = &      \frac{1}{2 \pi}    \nonumber  \\
\label{2-PhotonPhaseProbs}
\end{eqnarray}
\end{widetext}

In the limit of high photon absorption, $r_x = 1$ and $r_y =1$, the phase probability densities $p(\phi | m \, n, \, 2 0 ) = 1/(2 \pi)$ for all measurement outcomes, so that different  phase values are  not distinguishable because of photon losses.  

A remarkable property of the phase probability distributions in Eq.~(\ref{2-PhotonPhaseProbs}) (similar to that of Fock state input in Eq.~(\ref{1-PhotonPhaseDistributions}))  is that for equal losses in both arms of the interferometer, $r_x=r_y$, the phase probability distributions are independent of the magnitude of the loss $r_x$, having values
\begin{equation} 
\tiny
 \frac{1}{\pi }\left\{ {\frac{4}{3}\sin ^4 \left( {\frac{\phi }{2}} \right),\frac{4}{3}\cos ^4 \left( {\frac{\phi }{2}} \right),\sin ^2 (\phi ),\sin ^2 \left( {\frac{\phi }{2}} \right),\cos ^2 \left( {\frac{\phi }{2}} \right),\,\frac{1}{2}} \right\} 
\label{Fock2_rx=ry}
 \end{equation}
where I have written the values of the functions (from top to bottom) in Eq.~(\ref{2-PhotonPhaseProbs}) as components of a vector left to right in Eq.~(\ref{Fock2_rx=ry}).
%
%
%
%
One may think that for $r_x = r_y$  the contrast in the photon number measurements is lost, however, setting $r_x=r_y$ in Eq.~(\ref{2-PhotonMeasurementProb}) and expanding for small $r_x \ll 1$ leads to
\begin{equation}
\left( {\begin{array}{*{20}c}
   {P(20|\phi ,20)}  \\
   {P(02|\phi ,20)}  \\
   {P(11|\phi ,20)}  \\
   {P(10|\phi ,20)}  \\
   {P(01|\phi ,20)}  \\
   {P(00|\phi ,20)}  \\
\end{array}} \right) = \left( {\begin{array}{*{20}c}
   {\sin ^4 \left( {\frac{\phi }{2}} \right) - 2r_x^2 \sin ^4 \left( {\frac{\phi }{2}} \right) + O\left( {r_x^4 } \right)}  \\
   {\cos ^4 \left( {\frac{\phi }{2}} \right) - 2r_x^2 \cos ^4 \left( {\frac{\phi }{2}} \right) + O\left( {r_x^4 } \right)}  \\
   {\frac{{\sin ^2 (\phi )}}{2} - r_x^2 \sin ^2 (\phi ) + O\left( {r_x^4 } \right)}  \\
   {r_x^2 (1 - \cos (\phi )) + O\left( {r_x^4 } \right)}  \\
   {r_x^2 (\cos (\phi ) + 1) + O\left( {r_x^4 } \right)}  \\
   {O\left( {r_x^4 } \right)}  \\
\end{array}} \right)
\label{2-PhotonProbsSeriesExpansion}
\end{equation}
which shows that the probabilities become vanishingly small for the measurement outcomes $P(10|\phi ,20)$, $P(01|\phi ,20)$, and $P(00|\phi ,20)$, which correspond to one or both photons being absorbed, but contrast for different measurement outcomes is not lost. 

For the case  of equal loss in both arms, $r_x=r_y$, with no prior information on the phase, $p(\phi)= 1/(2 \pi)$, the fidelity is given by 

\begin{equation}
\small
H(M) = \frac{1}{{4\ln 2}}\left( {1 - r_x^2 } \right)     \left[ {8 - 4\ln 2 - 3\ln 3 + 2r_x^2 {\rm{arctanh}}\left( {\frac{{11}}{{43}}} \right)} \right]
\label{H-2PhotonFockRx=Ry}
\end{equation}

For the case of small (but not equal) losses in both arms, $r_x \ll 1$ and $r_y \ll 1$, the fidelity is given by

\begin{eqnarray}
H(M) = & &  \frac{{8 - 4\ln 2 - 3\ln 3}}{{4\ln 2}} + 
\left( {r_x^2  + r_y^2 } \right)\left( {\frac{{3\ln 3}}{{4\ln 2}} - \frac{1}{{\ln 2}}} \right) + \nonumber\\
 & & r_x^2 r_y^2 \left( {\frac{{1 + \ln 2 - \ln 3}}{{2\ln 2}}} \right) + O\left( {r_x } \right)^4  + O\left( {r_y } \right)^4 
\label{H-2PhotonFockRxRySmall}
\end{eqnarray}
where I have dropped terms of fourth order in $r_x$ and $r_y$.  Equation~(\ref{H-2PhotonFockRxRySmall}) gives the dependence on the dissipation parameters, $r_x$ and $r_y$, of the information gain about $\phi$, for single use of the interferometer, when there is no prior information about $\phi$.

\subsection{\label{N00NStateInputInput}N00N State Input into Lossy MZ Interferometer}
\begin{figure}[t]   
\includegraphics[width=3.1in]{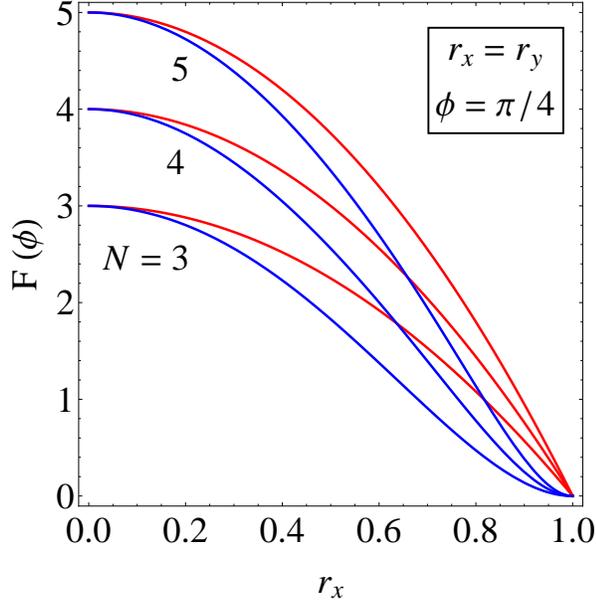}
\caption{\label{fig:Fisher_Comparison_Fock_N00N_states} (Color) The Fisher information is plotted vs. $r_x$ for $r_x=r_y$, for Fock states (red) and N00N states (blue), for $\phi=\pi/4$, where the Fisher information is non-zero for both Fock states and N00N states.}
\end{figure}

Next, I consider the $N$-photon N00N state input into the lossy Mach-Zehnder interferometer with scattering matrix given by Eq.~(\ref{LossyMZScatteringMatrix}):
\begin{eqnarray}
 \left| {\psi _{N00N} } \right\rangle  &  = & \frac{1}{{\sqrt {2N!} }}\left[ {\left( {\hat a_1^\dag  } \right)^N  + \left( {\hat a_2^\dag  } \right)^N } \right]\left| 0 \right\rangle   \\ 
          & = & \frac{1}{{\sqrt 2 }}\left[ {\left| {N000} \right\rangle  + \left| {0N00} \right\rangle } \right] 
\label{xxyyy}
\end{eqnarray}
For the input state in Eq.~(\ref{xxyyy}), the probability for measurement outcome $\xi= (n , m)$ is given by
\begin{eqnarray}
 & & P(n , m|\phi ,\psi _{N00N} )    =     
 \frac{N!}{2n!m!}    \sum\limits_{k = 0}^N \sum\limits_{l = 0}^N   \frac{1}{k!l!}  \times  \nonumber \\ 
 & &  \left| S_{11}^n S_{21}^m S_{31}^k S_{41}^l  + S_{12}^n S_{22}^m S_{32}^k S_{42}^l  \right|^2 \,\delta _{n + m + k + l,N}   
\label{MeasurementProbsForN00N} 
\end{eqnarray}
where $n$ and $m$ are the number of photons output in ports $b_1$ and $b_2$, respectively.  Using the Fisher information, I compare how well Fock states and N00N states perform  in the presence of absorption losses.  In Fig.~\ref{fig:Fisher_Comparison_Fock_N00N_states}, I plot the classical Fisher information for $N=$3, 4, and 5 photon Fock states and N00N states, plotted vs. $r_x$ for the special case where $r_x=r_y$. The plots show that, for equal dissipation in both arms, and for equal photon number, Fock states perform better for phase estimation than N00N states, for the same amount of dissipation $r_x$, see Eq.~(\ref{Cramer-RaoBound}). 

For N00N states, the Fisher information vanishes at the phase values: $\phi=0, \pm \pi/2, \pm \pi$.  While for Fock states, the Fisher information vanished only at $\phi=0, \pm \pi$.  For $\phi$ close to these values, phase estimation may have large standard deviation, see Eq.(\ref{Cramer-RaoBound}).   

\begin{figure}[t]   
\includegraphics[width=3.4in]{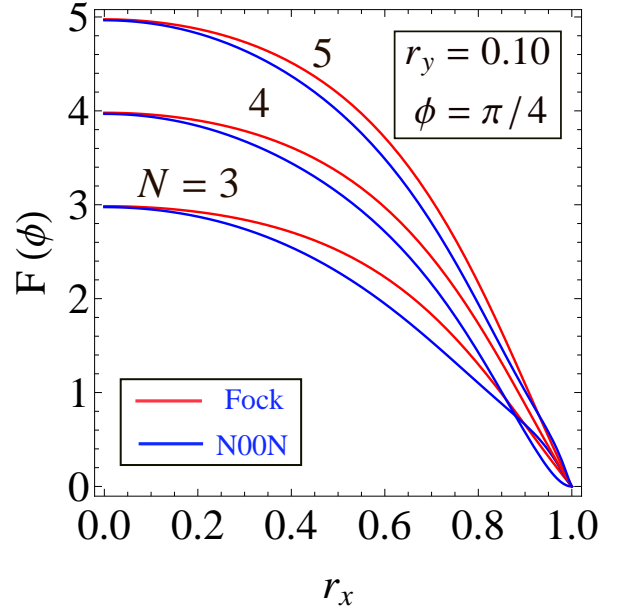}
\caption{\label{fig:Fisher_Comparison_SmallDissipation_Fock_N00N_states} (Color) The Fisher information is plotted vs. $r_x$ for small dissipation in one arm, $r_y=0.10$, for Fock states (red) and N00N states (blue), for $\phi=\pi/4$.}
\end{figure}

Figure~\ref{fig:Fisher_Comparison_SmallDissipation_Fock_N00N_states} shows the classical Fisher information for the case where Fock and N00N states are input into a Mach-Zehnder interferometer with small dissipation (losses) and when the losses are not equal in both arms. Generally, Fock states perform better (have larger Fisher information) for all values of dissipation $r_x$ except at the very highest values of $r_x\sim 0.95$.  The comparison is made at a true value of $\phi=\pi/4$, where the Fisher informations do not vanish.

\begin{figure}[t]   
\includegraphics[width=3.6in]{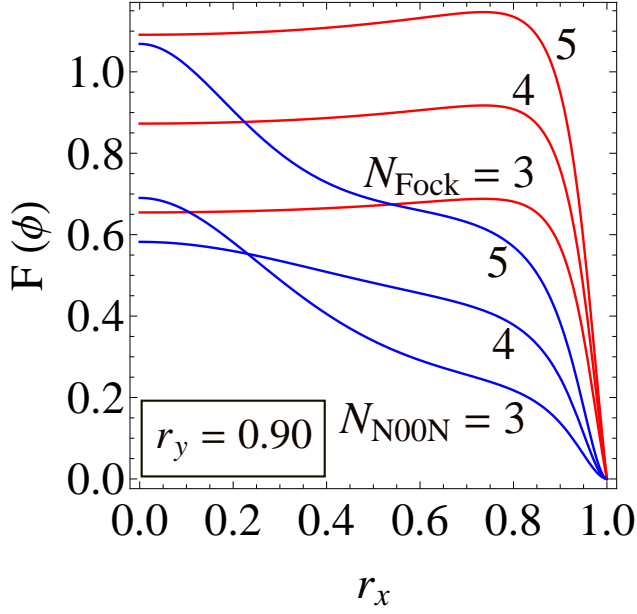}
\caption{\label{fig:Fisher_Comparison_LargeDissipation_Fock_N00N_states} (Color) The Fisher information is plotted vs. $r_x$ for large dissipation in one arm, $r_y=0.90$, for Fock states (red) and N00N states (blue), for $\phi=\pi/4$.  This example of high dissipation shows that the situation is complicated at high values of the dissipation in one arm and small values of dissipation in the other arm.}
\end{figure}

Figure~\ref{fig:Fisher_Comparison_LargeDissipation_Fock_N00N_states} shows a plot of the classical Fisher information for Fock and for N00N states for $N=$ 3, 4 and 5 photons for the case of large dissipation in one arm of the Mach-Zehnder interferometer, $r_y=0.9$.  The comparison is complicated, since for the $N=$3 photon case, Fock states perform better than N00N states for large dissipation $r_x \sim 0.8$, whereas the situation is reversed for small dissipation $r_x \sim 0.05$.   

\begin{figure}[t]   
\includegraphics[width=3.5in]{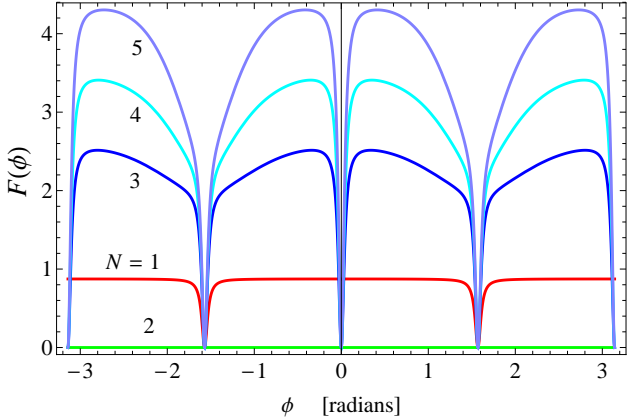}
\caption{\label{fig:Fisher_N00N_states} (Color) The Fisher information is plotted vs. $\phi$ for loss parameters $r_x=0.3$ and $r_y=0.4$, for N00N state input into a Mach-Zehnder interferometer for $N=$1, 2, 3, 4, and 5 photons.}
\end{figure}

In Figures~\ref{fig:Fisher_Comparison_SmallDissipation_Fock_N00N_states} and  \ref{fig:Fisher_Comparison_LargeDissipation_Fock_N00N_states}, the comparisons are made at a true value of phase $\phi=\pi/4$, where the classical Fisher informations (for Fock and N00N states) do not vanish.  When dissipation is present, the classical Fisher information has a complicated behavior as a function of the true phase $\phi$, see Fig.~\ref{fig:Fisher_N00N_states}.  This shows that phase estimation using simple photon counting is a sensitive procedure whose accuracy depends on the true value of phase.

\begin{figure}[t]   
\includegraphics[width=3.4in]{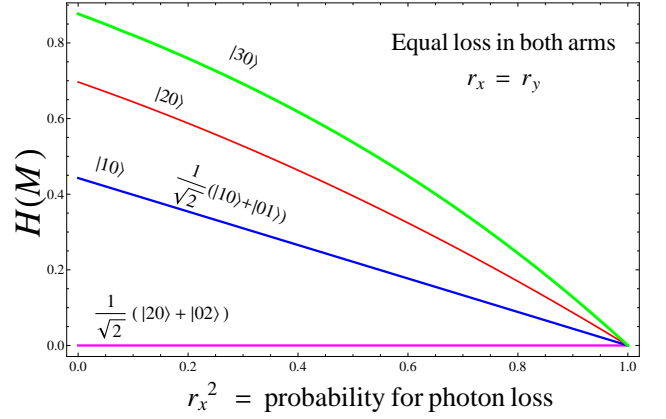}
\caption{\label{fig:FidelityComparison} (Color) The fidelity (Shannon mutual information) is plotted as a function of dissipation $r_x$, for $r_x=r_y$, for one-, two- and three-photon Fock state input, and for one- and  two-photon N00N state input.}
\end{figure}

The classical Fisher information for a lossy Mach-Zehnder interferometer depends on the true value of the phase $\phi$.  The fidelity (Shannon mutual information) is an information measure that averages over all phases, for prior information given by $p(\phi)$, see Eq.~(\ref{ShannonMutualInformation}).  Figure~\ref{fig:FidelityComparison} shows a comparison of the fidelity versus dissipation $r_x$ for Fock and N00N states for equal dissipation in both arms, $r_x = r_y$.   The fidelity of 1-photon Fock and N00N states is equal, see the discussion below.   For a given amount of dissipation, $r_x$, for Fock states the fidelity increases with input photon number $N$.  The fidelity for 2-photon N00N state input is exactly zero for all values of dissipation $r_x$ because this state carries no information about the phase in a Mach-Zehnder interferometer, see the discussion below.

\subsubsection{\label{1-PhotonN00NInput}1-Photon N00N State Input into Lossy MZ Interferometer}

Consider now the 1-photon entangled N00N state: 
\begin{equation}
\left| {\psi _{1}^{N00N} } \right\rangle  = \frac{1}{{\sqrt 2 }}\left( {a_1^\dag   + a_2^\dag  } \right)\left| 0 \right\rangle
= \frac{1}{\sqrt{2}}  \left[   \left| 1 0 \right\rangle  +    \left| 0 1  \right\rangle \right] 
\label{1-photon-N00N-state}
\end{equation}
where again, I use the short-hand notation $\left| 1 0 \right\rangle$ for $\left| 1 0 0 0 \right\rangle$ and $\left| 0 1 \right\rangle$ for $\left| 0 1 0 0 \right\rangle$.
The probabilities for the measurement outcomes for this input state are given by Eq.~(\ref{1-PhotonOutcome}) with the replacement $\cos \phi \rightarrow \sin \phi$.  Similarly, the phase probability distributions, assuming no prior information, $p(\phi)=1/(2 \pi)$,  are given by Eq.~(\ref{1-PhotonPhaseDistributions}) with the replacement $\cos \phi \rightarrow \sin \phi$.  The fidelity  for this input state is the same as for the 1-photon Fock state, given by Eq.~(\ref{1-PhotonFidelity-rx=ry=0})--(\ref{1-PhotonFidelitySmallRxRy}).   Therefore, according to Shannon mutual information (fidelity), the presence of entanglement in the 1-photon N00N state has not improved the information on the phase.  

The Fisher information for this entangled state is given by the 1-photon Fock state Fisher information in Eq.~(\ref{Fisher-1-Photon}) with the replacements $\phi \rightarrow \frac{\pi}{2} -\phi$. Therefore, the entanglement simply has the effect of changing the phase of the classical Fisher information. This phase change changes the places where $F(\phi)=0$, which, for this entangled state, is now $\phi= \pm \pi/2$.  Comparison of the 1-photon Fock state to the the 1-photon entangled N00N state shows that the introduction of entanglement does not remove the $\phi$ dependence of the Fisher information when arbitrary losses $r_x$ and $r_y$ are present.     
However, when $r_x = r_y$", the Fisher information, $F_1(\pi/2 - \phi)$, is independent of $\phi$. This  is in agreement with the result of Chen and Jiang, who derived the Fisher information for N00N state input using a master equation  for a quantum continuous variable system for the case of symmetrical losses~\cite{Chen2007}. 
When losses are absent, $r_x=r_y=0$, the Fisher information for input state given by Eq.~(\ref{1-photon-N00N-state}) reduces to $F(\phi)=1$, independent of $\phi$, as in the 1-photon Fock state without losses.    

\subsubsection{\label{2-PhotonN00NInput}2-Photon N00N State Input into Lossy MZ Interferometer}
The 2-photon N00N state,
\begin{equation}
\left| {\psi _2^{N00N} } \right\rangle  = \frac{1}{2}\left[ {\left( {\hat a_1^\dag  } \right)^2  + \left( {\hat a_2^\dag  } \right)^2 } \right]\,\left| 0 \right\rangle  =  \frac{1}{\sqrt{2}} \left[  \left| 2 0 \ \right \rangle +\left| 0 2 \ \right \rangle  \right] 
\label{2-photon-N00NState}
\end{equation}
has a peculiar behavior when input into a Mach-Zehnder interferometer with losses. The probabilities distributions, given by Eqs.~(\ref{MeasProbs})  and (\ref{MeasurementProbsForN00N}), for the six measurement outcomes are independent of $\phi$ and are given by
\begin{eqnarray}
P(20|\phi ,20) & = & P(02|\phi ,20) = \frac{1}{2} \left( 1 - r_x^2  - r_y^2  + r_x^2 r_y^2 \right) \nonumber \\
P(11|\phi ,20) & = & 0    \nonumber \\
P(10|\phi ,20) & = & P(01|\phi ,20) = \frac{1}{2} \left( r_x^2  + r_y^2 -2 r_x^2 r_y^2 \right)   \nonumber \\
P(00|\phi ,20) & = & r_x^2 r_y^2    \nonumber \\
\label{2-Photon_N00NStateProbs}
\end{eqnarray} 
The measurement outcomes in Eq.(66) are independent of $\phi$ because of the Hilbert space geometry of the measurement operators, $\hat{\Pi}_\phi (n,m)$, and input state vector in Eq.(\ref{2-photon-N00NState}), 
see Eq.(\ref{MeasurementOutcomes2}).  For no prior information on the phase, $p(\phi)=1/(2 \pi)$, using Bayes' rule in Eq.~(\ref{PhaseProbabilityDistribution}), the phase probability densities are independent of $\phi$, and are given by  $p(\phi |m \,n,20) = 1/(2 \pi)$, for all measurement outcomes $\xi=(m , n)$.  Therefore, the 2-photon N00N state cannot be used in a Mach-Zehnder interferometer for determining the phase $\phi$.   However, such an arrangement can be useful in applications that require phase in-sensitive interferometry to be performed.  In the limit of no loss, $r_x = r_y =0$, the interferometer acts as a beam splitter and both photons come out the same port. 
\begin{figure}[t]   
\includegraphics[width=3.4in]{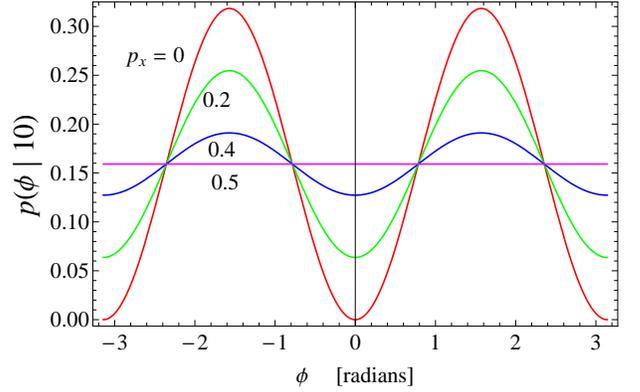}
\caption{\label{fig:ImperfectDetectorPhaseProbabilities} (Color) The conditional phase probability density, $p(\phi|10)$, is plotted as a function of $\phi$ for increasing detector error probabilities $p_x=0, \, 0.2, \, 0.4, \, 0.5$, showing a loss of phase distinguishability.  At $p_x=0.5$, all phases $\phi$ are equally probable.} 
\end{figure}
\begin{figure}[t]   
\includegraphics[width=3.2in]{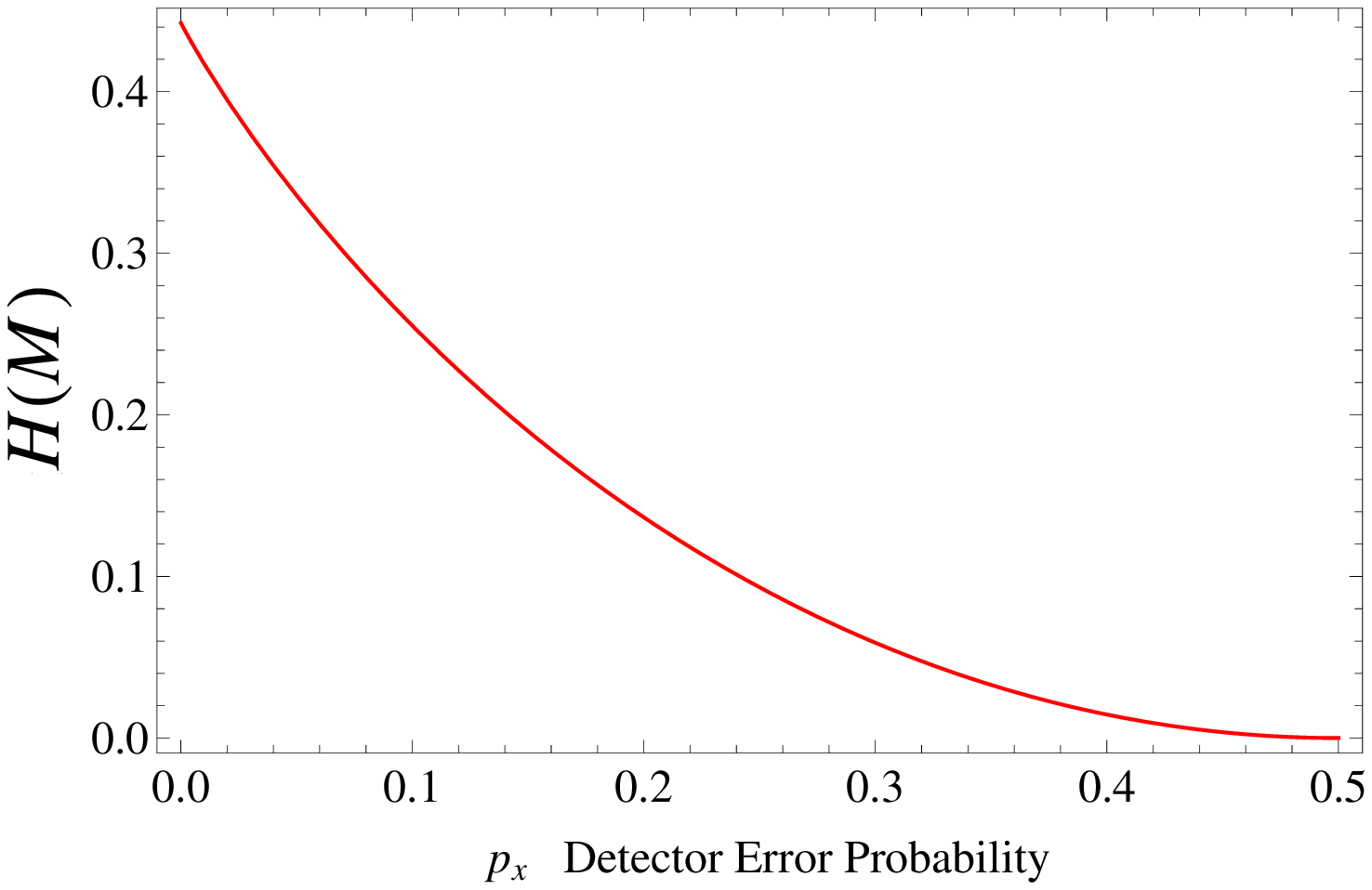}
\caption{\label{fig:ImperfectDetectorFidelity-1-photon} (Color) The fidelity (Shannon mutual information) between the measurements and the phase is plotted vs. the  probability of incorrect detection, $p_x$. Consistent with the phase probability density plotted in Fig.~\ref{fig:ImperfectDetectorPhaseProbabilities}, the fidelity decreases to zero at $p_x=0.5$ because there is no information on the phase in the measurements, so there is no discrimination between different phases.}
\end{figure}

\subsection{\label{ImperfectPhotonNumberDetection}Imperfect Photon Number Detection}

Next, I consider the simplest example of imperfect photon-number detection.  I assume that state preparation is deterministic and that the  interferometer is ideal, so there are no losses.   I also assume that the input state is a pure state, in Eq.~(\ref{MeasurementProbability}) taking  
\begin{equation}
P_S (\psi^{in} ) = \left\{ \begin{array}{l}
 1, \quad {\rm{if}}\;\left| {\psi^{in} } \right\rangle  = \left| {10} \right\rangle  \\ 
 0, \quad {\rm{otherwise}} \\ 
 \end{array} \right.
\label{PSdef-1-photon}
\end{equation}
where again I use the short-hand notation $ \left| {10} \right\rangle$ for $\left| 1000 \right\rangle  $.
For this 1-photon input state, $\left| {\psi^{in} } \right\rangle  = \left| {10} \right\rangle$,  the no-loss Mach-Zehnder interferometer transfer matrix is given by
\begin{equation}
\small
P_I (\psi^{out} |\psi^{in} ,\phi ) = \left\{ \begin{array}{l}
 \sin ^2 \phi ,\;{\rm{for}}\;\left| {\psi^{in} } \right\rangle  = \left| {10} \right\rangle {\rm{and}}\;\left| {\psi^{out} } \right\rangle  = \left| {10} \right\rangle  \\ 
 \cos ^2 \phi ,\;{\rm{for}}\;\left| {\psi^{in} } \right\rangle  = \left| {10} \right\rangle {\rm{and}}\;\left| {\psi^{out} } \right\rangle  = \left| {01} \right\rangle  \\ 
 0,  \quad {\rm{otherwise}} \\ 
 \end{array} \right.
\label{PI-Transfermatrix}
\end{equation}
For the  detection system, I assume that there is a probability $p_d$ to detect the state correctly and a probability $p_x$ to detect the state  incorrectly, where $p_d + p_x = 1$.  I am neglecting the possibility of an inconclusive measurement outcome.  The matrix, $P_D (\xi |\psi^{out} ,\phi )$, in Eq.~(\ref{MeasurementProbability}) describing the state detection is then given by
\begin{equation}
P_D (  \xi  |\psi^{out} ,\phi ) = \left\{ {\begin{array}{*{20}c}
   {p_d ,\quad \xi  = \psi^{out} }  \\
   {p_x ,\quad \xi  \ne \psi^{out} }  \\
\end{array}} \right.
\label{PD-1-photon}
\end{equation}
From  Eq.~(\ref{MeasurementProbability}),  the probabilities $P(\xi |\phi)$  for measurement outcomes are
\begin{equation}
\begin{array}{l}
 P(10|\phi ) = p_d \sin ^2 \phi  + p_x \cos ^2 \phi  \\ 
 P(01|\phi ) = p_x \sin ^2 \phi  + p_d \cos ^2 \phi  \\ 
 \end{array}
\label{DetectorProbs-1-photon}
\end{equation}
where $\xi=(m,n)$ specifies that $m$ and $n$ photons are detected in output ports ``c" and ``d", respectively, see Eq.~(\ref{InputOutputDef})--(\ref{MeasurementProbabilityDensitymatrix2}).

From Bayes' rule in Eq.~(\ref{PhaseProbabilityDistribution}), I find the
conditional probability density, $p(\phi | \xi )$, for the phase shift $\phi$
for a given measurement outcome $\xi$ to be
\begin{equation}
\begin{array}{l}
 p(\phi |10) = \frac{1}{\pi }\left[ {\left( {1 - p_x } \right)\sin ^2 \phi  + p_x \cos ^2 \phi } \right] \\ 
 p(\phi |01) = \frac{1}{\pi }\left[ {p_x \sin ^2 \phi  + \left( {1 - p_x } \right)\cos ^2 \phi } \right] \\ 
 \end{array}
\label{DetectorPhaseProbs}
\end{equation}
Figure~\ref{fig:ImperfectDetectorPhaseProbabilities} shows a plot of the phase probability density, $p(\phi | 10)$ vs. $\phi$,  for different values of detector error probability $p_x$.   
With increasing probability $p_x$ of detecting the state incorrectly, the constrast in the phase probability decreases.  Note that this contrast is not a ``visibility" because $p(\phi | 10)$ is a probability, and not an optical intensity. 

The fidelity (Shannon mutual information) defined in Eq.~(\ref{ShannonMutualInformation}) is plotted in Fig.~\ref{fig:ImperfectDetectorFidelity-1-photon}.  As expected for this simple model, the fidelity $H(M)$ decreases with increasing probability of incorrect detection, $p_x$, reaching zero at $p_x=0.5$. Note that the fidelity is symmetric about $p_x=0.5$.

\section{\label{Conclusion}Conclusion}
I considered the experimentally relevant problem of determining the phase shift in one arm of a quantum interferometer when state creation is not perfectly deterministic, state propagation through the interferometer is non-unitary due to absorption losses in the interferometer, and state detection is not ideal.     In Section II, I have argued that two types of information are useful for evaluating the quality of a parameter estimation device, such as a quantum optical system used to determine phase shifts.   First, fidelity (Shannon mutual information between measurements and parameter)  is useful for deciding the overall quality of the optical system. The fidelity represents an average over probabilities of all possible measurements and parameter values (phases).  The fidelity is the metric to use when choosing  or designing  a system and the prior parameter (phase) distribution is unknown.  Once a system  is chosen, it is to be used in estimating the parameter based on measurements (data), which is an estimation problem.  At this point, the (classical or quantum) Fisher information can be exploited,  using the classical or quantum Cramer-Rao theorem, to estimate the variance of the parameter associated with its unbiased estimator.  

In Eq.~(\ref{MeasurementProbability}), I have written down a general statistical expression for the probability of a measurement outcome that simultaneously takes into account the three non-ideal aspects of real experiments: non-deterministic state preparation, losses in the interferometer, and non-ideal quantum state detection.   This expression requires detailed models for each of the three non-ideal elements. 
In Section III, using simple, few-photon Fock states and N00N states, I give examples of applying Eq.~(\ref{MeasurementProbability}).  In subsection, A, of Section III, I look at a simple example of the effect of non-deterministic state creation, where there is a probability of creating one photon and a probability of creating vacuum, as input into a Mach-Zehnder interferometer. As expected, the non-zero probability of creating a vacuum input leads to a probability for an inconclusive measurement outcome, which in turn reduces the information on phase, as measured by Fisher information and fidelity (Shannon mutual information).

In subsection B, of Section III, I have constructed a scattering matrix for a lossy (non-unitary) Mach-Zehnder interferometer. I find that for simple photon counting measurements, losses introduce a strong phase dependence in the classical Fisher information, making accuracy of phase estimation dependent on the unknown true phase.  In subsection, C and D, of Section III, I use the classical Fisher information and fidelity to examine in detail the propagation of Fock and N00N states, respectively, through the lossy Mach-Zehnder interferometer. Finally, in subsection E of Section III, I use Eq.~(\ref{MeasurementProbability}), to look at the effect of imperfect photon number detection on determining the phase in a Mach-Zehnder interferometer with no losses, using the simplest model that takes into account a probability for incorrect detection of photon number.   

The examples that I have used have been simple to illustrate the application of Eq.~(\ref{MeasurementProbability}).  The theory can be applied to more complicated cases so that real experiments can be analyzed.

\begin{acknowledgments}
The author acknowledges stimulating discussions with Paul Lopata on Bayesian statistics and data analysis.  
\end{acknowledgments}
%
\bibliographystyle{apsrev}
\bibliography{References-Quantum}

\begin{thebibliography}{60}
\expandafter\ifx\csname natexlab\endcsname\relax\def\natexlab#1{#1}\fi
\expandafter\ifx\csname bibnamefont\endcsname\relax
  \def\bibnamefont#1{#1}\fi
\expandafter\ifx\csname bibfnamefont\endcsname\relax
  \def\bibfnamefont#1{#1}\fi
\expandafter\ifx\csname citenamefont\endcsname\relax
  \def\citenamefont#1{#1}\fi
\expandafter\ifx\csname url\endcsname\relax
  \def\url#1{\texttt{#1}}\fi
\expandafter\ifx\csname urlprefix\endcsname\relax\def\urlprefix{URL }\fi
\providecommand{\bibinfo}[2]{#2}
\providecommand{\eprint}[2][]{\url{#2}}

\bibitem[{\citenamefont{Hariharan}(2003)}]{Hariharan2003}
\bibinfo{author}{\bibfnamefont{P.}~\bibnamefont{Hariharan}},
  \emph{\bibinfo{title}{Optical interferometry}} (\bibinfo{publisher}{Academic
  Press}, \bibinfo{address}{New York}, \bibinfo{year}{2003}),
  \bibinfo{edition}{second edition} ed.

\bibitem[{\citenamefont{Cronin et~al.}(2009)\citenamefont{Cronin,
  Schmiedmayer†, and Pritchard}}]{Cronin2009}
\bibinfo{author}{\bibfnamefont{A.~D.} \bibnamefont{Cronin}},
  \bibinfo{author}{\bibfnamefont{J.}~\bibnamefont{Schmiedmayer†}},
  \bibnamefont{and} \bibinfo{author}{\bibfnamefont{D.~E.}
  \bibnamefont{Pritchard}}, \bibinfo{journal}{Rev. Mod. Phys.}
  \textbf{\bibinfo{volume}{81}}, \bibinfo{pages}{1051} (\bibinfo{year}{2009}).

\bibitem[{\citenamefont{Thorne}(1980)}]{Thorne1980}
\bibinfo{author}{\bibfnamefont{K.}~\bibnamefont{Thorne}},
  \bibinfo{journal}{Rev. Mod. Phys.} \textbf{\bibinfo{volume}{52}},
  \bibinfo{pages}{285} (\bibinfo{year}{1980}).

\bibitem[{\citenamefont{Caves}(1981)}]{Caves1981}
\bibinfo{author}{\bibfnamefont{C.~M.} \bibnamefont{Caves}},
  \bibinfo{journal}{Phys. Rev. D} \textbf{\bibinfo{volume}{23}},
  \bibinfo{pages}{1693} (\bibinfo{year}{1981}).

\bibitem[{\citenamefont{Dimopoulos et~al.}(2008)\citenamefont{Dimopoulos,
  Graham, Hogan, and Kasevich}}]{Dimopoulos2008}
\bibinfo{author}{\bibfnamefont{S.}~\bibnamefont{Dimopoulos}},
  \bibinfo{author}{\bibfnamefont{P.~W.} \bibnamefont{Graham}},
  \bibinfo{author}{\bibfnamefont{J.~M.} \bibnamefont{Hogan}}, \bibnamefont{and}
  \bibinfo{author}{\bibfnamefont{M.~A.} \bibnamefont{Kasevich}},
  \bibinfo{journal}{Phys. Rev. D} \textbf{\bibinfo{volume}{78}},
  \bibinfo{pages}{042003} (\bibinfo{year}{2008}).

\bibitem[{\citenamefont{Lefevre}(1993)}]{Lefevre1983}
\bibinfo{author}{\bibfnamefont{H.}~\bibnamefont{Lefevre}},
  \emph{\bibinfo{title}{The fiber-optic gyroscope}} (\bibinfo{publisher}{Artech
  House}, \bibinfo{address}{Boston, USA}, \bibinfo{year}{1993}).

\bibitem[{\citenamefont{Sagnac}(1913{\natexlab{a}})}]{Sagnac1913a}
\bibinfo{author}{\bibfnamefont{G.}~\bibnamefont{Sagnac}},
  \bibinfo{journal}{Compt. Rend.} \textbf{\bibinfo{volume}{157}},
  \bibinfo{pages}{708} (\bibinfo{year}{1913}{\natexlab{a}}).

\bibitem[{\citenamefont{Sagnac}(1913{\natexlab{b}})}]{Sagnac1913b}
\bibinfo{author}{\bibfnamefont{G.}~\bibnamefont{Sagnac}},
  \bibinfo{journal}{Compt. Rend.} \textbf{\bibinfo{volume}{157}},
  \bibinfo{pages}{1410} (\bibinfo{year}{1913}{\natexlab{b}}).

\bibitem[{\citenamefont{Sagnac}(1914)}]{Sagnac1914}
\bibinfo{author}{\bibfnamefont{G.}~\bibnamefont{Sagnac}}, \bibinfo{journal}{J.
  Phys. Radium} \textbf{\bibinfo{volume}{5th Series 4}}, \bibinfo{pages}{177}
  (\bibinfo{year}{1914}).

\bibitem[{\citenamefont{Post}(1967)}]{Post1967}
\bibinfo{author}{\bibfnamefont{E.~J.} \bibnamefont{Post}},
  \bibinfo{journal}{Rev. Mod. Phys.} \textbf{\bibinfo{volume}{39}},
  \bibinfo{pages}{475} (\bibinfo{year}{1967}).

\bibitem[{\citenamefont{Chen et~al.}(2008)\citenamefont{Chen, Altepeter, and
  Kumar}}]{Chen2008}
\bibinfo{author}{\bibfnamefont{J.}~\bibnamefont{Chen}},
  \bibinfo{author}{\bibfnamefont{J.~B.} \bibnamefont{Altepeter}},
  \bibnamefont{and} \bibinfo{author}{\bibfnamefont{P.}~\bibnamefont{Kumar}},
  \bibinfo{journal}{New J. Phys.} \textbf{\bibinfo{volume}{10}},
  \bibinfo{pages}{123019} (\bibinfo{year}{2008}).

\bibitem[{\citenamefont{Bertocchi et~al.}(2006)\citenamefont{Bertocchi,
  Alibart, Ostrowsky, Tanzilli, and Baldi}}]{Bertocchi2006}
\bibinfo{author}{\bibfnamefont{G.}~\bibnamefont{Bertocchi}},
  \bibinfo{author}{\bibfnamefont{O.}~\bibnamefont{Alibart}},
  \bibinfo{author}{\bibfnamefont{D.~B.} \bibnamefont{Ostrowsky}},
  \bibinfo{author}{\bibfnamefont{S.}~\bibnamefont{Tanzilli}}, \bibnamefont{and}
  \bibinfo{author}{\bibfnamefont{P.}~\bibnamefont{Baldi}}, \bibinfo{journal}{J.
  Phys. B} \textbf{\bibinfo{volume}{39}}, \bibinfo{pages}{1011}
  (\bibinfo{year}{2006}).

\bibitem[{\citenamefont{Gupta et~al.}(2005)\citenamefont{Gupta, Murch, Moore,
  Purdy, and Stamper-Kurn}}]{Gupta2005}
\bibinfo{author}{\bibfnamefont{S.}~\bibnamefont{Gupta}},
  \bibinfo{author}{\bibfnamefont{K.~W.} \bibnamefont{Murch}},
  \bibinfo{author}{\bibfnamefont{K.~L.} \bibnamefont{Moore}},
  \bibinfo{author}{\bibfnamefont{T.~P.} \bibnamefont{Purdy}}, \bibnamefont{and}
  \bibinfo{author}{\bibfnamefont{D.~M.} \bibnamefont{Stamper-Kurn}},
  \bibinfo{journal}{Phys. Rev. Lett.} \textbf{\bibinfo{volume}{95}},
  \bibinfo{pages}{143201} (\bibinfo{year}{2005}).

\bibitem[{\citenamefont{Wang et~al.}(2005)\citenamefont{Wang, Anderson, Bright,
  Cornell, Diot, Kishimoto, Prentiss, Saravanan, Segal, and Wu}}]{Wang2005}
\bibinfo{author}{\bibfnamefont{Y.-J.} \bibnamefont{Wang}},
  \bibinfo{author}{\bibfnamefont{D.~Z.} \bibnamefont{Anderson}},
  \bibinfo{author}{\bibfnamefont{V.~M.} \bibnamefont{Bright}},
  \bibinfo{author}{\bibfnamefont{E.~A.} \bibnamefont{Cornell}},
  \bibinfo{author}{\bibfnamefont{Q.}~\bibnamefont{Diot}},
  \bibinfo{author}{\bibfnamefont{T.}~\bibnamefont{Kishimoto}},
  \bibinfo{author}{\bibfnamefont{M.}~\bibnamefont{Prentiss}},
  \bibinfo{author}{\bibfnamefont{R.~A.} \bibnamefont{Saravanan}},
  \bibinfo{author}{\bibfnamefont{S.~R.} \bibnamefont{Segal}}, \bibnamefont{and}
  \bibinfo{author}{\bibfnamefont{S.}~\bibnamefont{Wu}}, \bibinfo{journal}{Phys.
  Rev. Lett.} \textbf{\bibinfo{volume}{94}}, \bibinfo{pages}{090405}
  (\bibinfo{year}{2005}).

\bibitem[{\citenamefont{Tolstikhin et~al.}(2005)\citenamefont{Tolstikhin,
  Morishita, and Watanabe}}]{Tolstikhin2005}
\bibinfo{author}{\bibfnamefont{O.~I.} \bibnamefont{Tolstikhin}},
  \bibinfo{author}{\bibfnamefont{T.}~\bibnamefont{Morishita}},
  \bibnamefont{and} \bibinfo{author}{\bibfnamefont{S.}~\bibnamefont{Watanabe}},
  \bibinfo{journal}{Phys. Rev. A} \textbf{\bibinfo{volume}{72}},
  \bibinfo{pages}{051603(R)} (\bibinfo{year}{2005}).

\bibitem[{\citenamefont{Cooper et~al.}(2010)\citenamefont{Cooper, Hallwood, and
  Dunningham}}]{Cooper2010}
\bibinfo{author}{\bibfnamefont{J.~J.} \bibnamefont{Cooper}},
  \bibinfo{author}{\bibfnamefont{D.~W.} \bibnamefont{Hallwood}},
  \bibnamefont{and} \bibinfo{author}{\bibfnamefont{J.~A.}
  \bibnamefont{Dunningham}}, \bibinfo{journal}{Phys. Rev. A}
  \textbf{\bibinfo{volume}{81}}, \bibinfo{pages}{043624}
  (\bibinfo{year}{2010}).

\bibitem[{\citenamefont{Godun et~al.}(2001)\citenamefont{Godun, d'Arcy, Summy,
  and Burnett}}]{Godun2001}
\bibinfo{author}{\bibfnamefont{R.~M.} \bibnamefont{Godun}},
  \bibinfo{author}{\bibfnamefont{M.~B.} \bibnamefont{d'Arcy}},
  \bibinfo{author}{\bibfnamefont{G.~S.} \bibnamefont{Summy}}, \bibnamefont{and}
  \bibinfo{author}{\bibfnamefont{K.}~\bibnamefont{Burnett}},
  \bibinfo{journal}{Contemporary Physics} \textbf{\bibinfo{volume}{42}},
  \bibinfo{pages}{77} (\bibinfo{year}{2001}).

\bibitem[{\citenamefont{Giovannetti et~al.}(2006)\citenamefont{Giovannetti,
  Lloyd, and Maccone}}]{Giovannetti2006}
\bibinfo{author}{\bibfnamefont{V.}~\bibnamefont{Giovannetti}},
  \bibinfo{author}{\bibfnamefont{S.}~\bibnamefont{Lloyd}}, \bibnamefont{and}
  \bibinfo{author}{\bibfnamefont{L.}~\bibnamefont{Maccone}},
  \bibinfo{journal}{Phys. Rev. Lett.} \textbf{\bibinfo{volume}{96}},
  \bibinfo{pages}{010401} (\bibinfo{year}{2006}).

\bibitem[{\citenamefont{Berry et~al.}(2009)\citenamefont{Berry, Higgins,
  Bartlett, Mitchell, Pryde, and Wiseman}}]{Berry2009}
\bibinfo{author}{\bibfnamefont{D.~W.} \bibnamefont{Berry}},
  \bibinfo{author}{\bibfnamefont{B.~L.} \bibnamefont{Higgins}},
  \bibinfo{author}{\bibfnamefont{S.~D.} \bibnamefont{Bartlett}},
  \bibinfo{author}{\bibfnamefont{M.~W.} \bibnamefont{Mitchell}},
  \bibinfo{author}{\bibfnamefont{G.~J.} \bibnamefont{Pryde}}, \bibnamefont{and}
  \bibinfo{author}{\bibfnamefont{H.~M.} \bibnamefont{Wiseman}},
  \bibinfo{journal}{Phys. Rev. A} \textbf{\bibinfo{volume}{80}},
  \bibinfo{pages}{052114} (\bibinfo{year}{2009}).

\bibitem[{\citenamefont{Combes and Wiseman}(2005)}]{Combes2005}
\bibinfo{author}{\bibfnamefont{J.}~\bibnamefont{Combes}} \bibnamefont{and}
  \bibinfo{author}{\bibfnamefont{H.~M.} \bibnamefont{Wiseman}},
  \bibinfo{journal}{J. Opt. B: Quantum Semiclass} \textbf{\bibinfo{volume}{7}},
  \bibinfo{pages}{14} (\bibinfo{year}{2005}).

\bibitem[{\citenamefont{Nagata et~al.}(2007)\citenamefont{Nagata, Okamoto,
  O'Brien, Sasaki, and Takeuchi}}]{Nagata2007}
\bibinfo{author}{\bibfnamefont{T.}~\bibnamefont{Nagata}},
  \bibinfo{author}{\bibfnamefont{R.}~\bibnamefont{Okamoto}},
  \bibinfo{author}{\bibfnamefont{J.~L.} \bibnamefont{O'Brien}},
  \bibinfo{author}{\bibfnamefont{K.}~\bibnamefont{Sasaki}}, \bibnamefont{and}
  \bibinfo{author}{\bibfnamefont{S.}~\bibnamefont{Takeuchi}},
  \bibinfo{journal}{Science} \textbf{\bibinfo{volume}{316}},
  \bibinfo{pages}{726} (\bibinfo{year}{2007}).

\bibitem[{\citenamefont{Durkin and Dowling}(2007)}]{Durkin2007}
\bibinfo{author}{\bibfnamefont{G.~A.} \bibnamefont{Durkin}} \bibnamefont{and}
  \bibinfo{author}{\bibfnamefont{J.~P.} \bibnamefont{Dowling}},
  \bibinfo{journal}{Phys. Rev. Lett.} \textbf{\bibinfo{volume}{99}},
  \bibinfo{pages}{070801} (\bibinfo{year}{2007}).

\bibitem[{\citenamefont{Pezze and Smerzi}(2008)}]{Pezze2008}
\bibinfo{author}{\bibfnamefont{L.}~\bibnamefont{Pezze}} \bibnamefont{and}
  \bibinfo{author}{\bibfnamefont{A.}~\bibnamefont{Smerzi}},
  \bibinfo{journal}{Phys. Rev. Lett.} \textbf{\bibinfo{volume}{100}},
  \bibinfo{pages}{073601} (\bibinfo{year}{2008}).

\bibitem[{\citenamefont{Dorner et~al.}(2009)\citenamefont{Dorner,
  Demkowicz-Dobrzanski, Smith, Lundeen, Wasilewski, Banaszek, and
  Walmsley}}]{Dorner2009}
\bibinfo{author}{\bibfnamefont{U.}~\bibnamefont{Dorner}},
  \bibinfo{author}{\bibfnamefont{R.}~\bibnamefont{Demkowicz-Dobrzanski}},
  \bibinfo{author}{\bibfnamefont{B.~J.} \bibnamefont{Smith}},
  \bibinfo{author}{\bibfnamefont{J.~S.} \bibnamefont{Lundeen}},
  \bibinfo{author}{\bibfnamefont{W.}~\bibnamefont{Wasilewski}},
  \bibinfo{author}{\bibfnamefont{K.}~\bibnamefont{Banaszek}}, \bibnamefont{and}
  \bibinfo{author}{\bibfnamefont{I.~A.} \bibnamefont{Walmsley}},
  \bibinfo{journal}{Phys. Rev. Lett.} \textbf{\bibinfo{volume}{102}},
  \bibinfo{pages}{040403} (\bibinfo{year}{2009}).

\bibitem[{\citenamefont{Cable and Durkin}(2010)}]{Cable2010}
\bibinfo{author}{\bibfnamefont{H.}~\bibnamefont{Cable}} \bibnamefont{and}
  \bibinfo{author}{\bibfnamefont{G.~A.} \bibnamefont{Durkin}},
  \bibinfo{journal}{Phys. Rev. Lett.} \textbf{\bibinfo{volume}{105}},
  \bibinfo{pages}{013603} (\bibinfo{year}{2010}).

\bibitem[{\citenamefont{Bahder and Lopata}(2006{\natexlab{a}})}]{Bahder2006}
\bibinfo{author}{\bibfnamefont{T.~B.} \bibnamefont{Bahder}} \bibnamefont{and}
  \bibinfo{author}{\bibfnamefont{P.~A.} \bibnamefont{Lopata}},
  \bibinfo{journal}{Phys. Rev. A} \textbf{\bibinfo{volume}{74}},
  \bibinfo{pages}{051801R} (\bibinfo{year}{2006}{\natexlab{a}}),
  \urlprefix\url{http://arxiv.org/abs/quant-ph/0602123}.

\bibitem[{\citenamefont{Cram\'er}(1958)}]{Cramer1958}
\bibinfo{author}{\bibfnamefont{H.}~\bibnamefont{Cram\'er}},
  \emph{\bibinfo{title}{Mathematical Methods of Statistics}}
  (\bibinfo{publisher}{Princeton University Press},
  \bibinfo{address}{Princeton}, \bibinfo{year}{1958}), \bibinfo{note}{eighth
  printing}.

\bibitem[{\citenamefont{Helstrom}(1967)}]{Helstrom1967}
\bibinfo{author}{\bibfnamefont{C.~W.} \bibnamefont{Helstrom}},
  \bibinfo{journal}{Phys. Lett. A} \textbf{\bibinfo{volume}{25}},
  \bibinfo{pages}{101} (\bibinfo{year}{1967}).

\bibitem[{\citenamefont{Helstrom}(1976)}]{Helstrom1976}
\bibinfo{author}{\bibfnamefont{C.~W.} \bibnamefont{Helstrom}},
  \emph{\bibinfo{title}{Quantum Detection and Estimation Theory}}
  (\bibinfo{publisher}{Academic Press}, \bibinfo{address}{New York},
  \bibinfo{year}{1976}).

\bibitem[{\citenamefont{Holevo}(1982)}]{Holevo1982}
\bibinfo{author}{\bibfnamefont{A.~S.} \bibnamefont{Holevo}},
  \emph{\bibinfo{title}{Probabilistic and Statistical Aspects of Quantum
  Theory}} (\bibinfo{publisher}{North-Holland}, \bibinfo{address}{Amsterdam},
  \bibinfo{year}{1982}).

\bibitem[{\citenamefont{Braunstein and Caves}(1994)}]{Braunstein1994}
\bibinfo{author}{\bibfnamefont{S.~L.} \bibnamefont{Braunstein}}
  \bibnamefont{and} \bibinfo{author}{\bibfnamefont{C.~M.} \bibnamefont{Caves}},
  \bibinfo{journal}{Phys. Rev. Lett.} \textbf{\bibinfo{volume}{72}},
  \bibinfo{pages}{3439} (\bibinfo{year}{1994}).

\bibitem[{\citenamefont{Braunstein et~al.}(1996)\citenamefont{Braunstein,
  Caves, and Milburn}}]{Braunstein1996}
\bibinfo{author}{\bibfnamefont{S.~L.} \bibnamefont{Braunstein}},
  \bibinfo{author}{\bibfnamefont{C.~M.} \bibnamefont{Caves}}, \bibnamefont{and}
  \bibinfo{author}{\bibfnamefont{G.~J.} \bibnamefont{Milburn}},
  \bibinfo{journal}{Ann. of Phys.} \textbf{\bibinfo{volume}{247}},
  \bibinfo{pages}{135} (\bibinfo{year}{1996}).

\bibitem[{\citenamefont{Barndorff-Nielsen and
  Gill}(2000)}]{Barndorff-Nielsen2000}
\bibinfo{author}{\bibfnamefont{O.~E.} \bibnamefont{Barndorff-Nielsen}}
  \bibnamefont{and} \bibinfo{author}{\bibfnamefont{R.~D.} \bibnamefont{Gill}},
  \bibinfo{journal}{J. Phys. A: Math. Gen.} \textbf{\bibinfo{volume}{33}},
  \bibinfo{pages}{4481} (\bibinfo{year}{2000}).

\bibitem[{\citenamefont{Barndorff-Nielsen
  et~al.}(2003)\citenamefont{Barndorff-Nielsen, Gill, and
  Jupp}}]{Barndorff-Nielsen2003}
\bibinfo{author}{\bibfnamefont{O.~E.} \bibnamefont{Barndorff-Nielsen}},
  \bibinfo{author}{\bibfnamefont{R.~D.} \bibnamefont{Gill}}, \bibnamefont{and}
  \bibinfo{author}{\bibfnamefont{P.~E.} \bibnamefont{Jupp}},
  \bibinfo{journal}{J. Roy. Stat. Soc. B} \textbf{\bibinfo{volume}{65}},
  \bibinfo{pages}{775} (\bibinfo{year}{2003}),
  \urlprefix\url{http://arxiv.org/abs/quant-ph/0307191}.

\bibitem[{\citenamefont{Walther et~al.}(2004)\citenamefont{Walther, Pan,
  Aspelmeyer, Ursin, Gasparoni, and Zeilinger}}]{Walther2004}
\bibinfo{author}{\bibfnamefont{P.}~\bibnamefont{Walther}},
  \bibinfo{author}{\bibfnamefont{J.}~\bibnamefont{Pan}},
  \bibinfo{author}{\bibfnamefont{M.}~\bibnamefont{Aspelmeyer}},
  \bibinfo{author}{\bibfnamefont{R.}~\bibnamefont{Ursin}},
  \bibinfo{author}{\bibfnamefont{S.}~\bibnamefont{Gasparoni}},
  \bibnamefont{and}
  \bibinfo{author}{\bibfnamefont{A.}~\bibnamefont{Zeilinger}},
  \bibinfo{journal}{Nature} \textbf{\bibinfo{volume}{429}},
  \bibinfo{pages}{158} (\bibinfo{year}{2004}).

\bibitem[{\citenamefont{Mitchell et~al.}(2004)\citenamefont{Mitchell, Lundeen,
  and Steinberg}}]{Mitchell2004}
\bibinfo{author}{\bibfnamefont{M.~W.} \bibnamefont{Mitchell}},
  \bibinfo{author}{\bibfnamefont{J.~S.} \bibnamefont{Lundeen}},
  \bibnamefont{and} \bibinfo{author}{\bibfnamefont{A.~M.}
  \bibnamefont{Steinberg}}, \bibinfo{journal}{Nature}
  \textbf{\bibinfo{volume}{429}}, \bibinfo{pages}{161} (\bibinfo{year}{2004}).

\bibitem[{\citenamefont{Okamoto et~al.}(2008)\citenamefont{Okamoto, Hofmann,
  Nagata, O'Brien, Sasaki, and Takeuchi}}]{Okamoto2008}
\bibinfo{author}{\bibfnamefont{R.}~\bibnamefont{Okamoto}},
  \bibinfo{author}{\bibfnamefont{H.~F.} \bibnamefont{Hofmann}},
  \bibinfo{author}{\bibfnamefont{T.}~\bibnamefont{Nagata}},
  \bibinfo{author}{\bibfnamefont{J.~L.} \bibnamefont{O'Brien}},
  \bibinfo{author}{\bibfnamefont{K.}~\bibnamefont{Sasaki}}, \bibnamefont{and}
  \bibinfo{author}{\bibfnamefont{S.}~\bibnamefont{Takeuchi}},
  \bibinfo{journal}{New J. Phys.} \textbf{\bibinfo{volume}{10}},
  \bibinfo{pages}{073033} (\bibinfo{year}{2008}).

\bibitem[{\citenamefont{Kacprowicz et~al.}(2010)\citenamefont{Kacprowicz,
  Demkowicz-Dobrzanski, Wasilewski, Banaszek, and Walmsley}}]{Kacprowicz2010}
\bibinfo{author}{\bibfnamefont{M.}~\bibnamefont{Kacprowicz}},
  \bibinfo{author}{\bibfnamefont{R.}~\bibnamefont{Demkowicz-Dobrzanski}},
  \bibinfo{author}{\bibfnamefont{W.}~\bibnamefont{Wasilewski}},
  \bibinfo{author}{\bibfnamefont{K.}~\bibnamefont{Banaszek}}, \bibnamefont{and}
  \bibinfo{author}{\bibfnamefont{I.~A.} \bibnamefont{Walmsley}},
  \bibinfo{journal}{Nature Photonics} \textbf{\bibinfo{volume}{4}},
  \bibinfo{pages}{357} (\bibinfo{year}{2010}),
  \urlprefix\url{http://lanl.arxiv.org/abs/0906.3511}.

\bibitem[{\citenamefont{Thomas-Peter et~al.}(2009)\citenamefont{Thomas-Peter,
  Smith, and Walmsley}}]{Thomas-Peter2009}
\bibinfo{author}{\bibfnamefont{N.}~\bibnamefont{Thomas-Peter}},
  \bibinfo{author}{\bibfnamefont{B.~J.} \bibnamefont{Smith}}, \bibnamefont{and}
  \bibinfo{author}{\bibfnamefont{I.~A.} \bibnamefont{Walmsley}}, in
  \emph{\bibinfo{booktitle}{Lasers and Electro-Optics, 2009 and 2009 Conference
  on Quantum electronics and Laser Science Conference. CLEO/QELS 2009.}}
  (\bibinfo{address}{Baltimore, MD}, \bibinfo{year}{2009}), pp.
  \bibinfo{pages}{978--1--55752--869--8}.

\bibitem[{\citenamefont{Kim et~al.}(1998)\citenamefont{Kim, Pfister, Holland,
  Noh, and Hall}}]{Kim1998}
\bibinfo{author}{\bibfnamefont{T.}~\bibnamefont{Kim}},
  \bibinfo{author}{\bibfnamefont{O.}~\bibnamefont{Pfister}},
  \bibinfo{author}{\bibfnamefont{M.~J.} \bibnamefont{Holland}},
  \bibinfo{author}{\bibfnamefont{J.}~\bibnamefont{Noh}}, \bibnamefont{and}
  \bibinfo{author}{\bibfnamefont{J.~L.} \bibnamefont{Hall}},
  \bibinfo{journal}{Phys. Rev. A} \textbf{\bibinfo{volume}{57}},
  \bibinfo{pages}{4004} (\bibinfo{year}{1998}).

\bibitem[{\citenamefont{Durkin et~al.}(2004)\citenamefont{Durkin, Simon,
  Eisert, and Bouwmeester}}]{Durkin2004}
\bibinfo{author}{\bibfnamefont{G.~A.} \bibnamefont{Durkin}},
  \bibinfo{author}{\bibfnamefont{C.}~\bibnamefont{Simon}},
  \bibinfo{author}{\bibfnamefont{J.}~\bibnamefont{Eisert}}, \bibnamefont{and}
  \bibinfo{author}{\bibfnamefont{D.}~\bibnamefont{Bouwmeester}},
  \bibinfo{journal}{Phys. Rev. A} \textbf{\bibinfo{volume}{70}},
  \bibinfo{pages}{062305} (\bibinfo{year}{2004}).

\bibitem[{\citenamefont{Rubin and Kaushik}(2007)}]{Rubin2007}
\bibinfo{author}{\bibfnamefont{M.~A.} \bibnamefont{Rubin}} \bibnamefont{and}
  \bibinfo{author}{\bibfnamefont{S.}~\bibnamefont{Kaushik}},
  \bibinfo{journal}{Phys. Rev. A} \textbf{\bibinfo{volume}{75}},
  \bibinfo{pages}{053805} (\bibinfo{year}{2007}).

\bibitem[{\citenamefont{Gilbert et~al.}(2008)\citenamefont{Gilbert, Hamrick,
  and Weinstein}}]{Gilbert2008}
\bibinfo{author}{\bibfnamefont{G.}~\bibnamefont{Gilbert}},
  \bibinfo{author}{\bibfnamefont{M.}~\bibnamefont{Hamrick}}, \bibnamefont{and}
  \bibinfo{author}{\bibfnamefont{Y.~S.} \bibnamefont{Weinstein}},
  \bibinfo{journal}{J. Opt. Soc. Am. B} \textbf{\bibinfo{volume}{25}},
  \bibinfo{pages}{1336} (\bibinfo{year}{2008}).

\bibitem[{\citenamefont{Demkowicz-Dobrzanski
  et~al.}(2009)\citenamefont{Demkowicz-Dobrzanski, Dorner, Smith, Lundeen,
  Wasilewski, Banaszek, and Walmsley}}]{Demkowicz-Dobrzanski2009}
\bibinfo{author}{\bibfnamefont{R.}~\bibnamefont{Demkowicz-Dobrzanski}},
  \bibinfo{author}{\bibfnamefont{U.}~\bibnamefont{Dorner}},
  \bibinfo{author}{\bibfnamefont{B.~J.} \bibnamefont{Smith}},
  \bibinfo{author}{\bibfnamefont{J.~S.} \bibnamefont{Lundeen}},
  \bibinfo{author}{\bibfnamefont{W.}~\bibnamefont{Wasilewski}},
  \bibinfo{author}{\bibfnamefont{K.}~\bibnamefont{Banaszek}}, \bibnamefont{and}
  \bibinfo{author}{\bibfnamefont{I.~A.} \bibnamefont{Walmsley}},
  \bibinfo{journal}{Phys. Rev. A} \textbf{\bibinfo{volume}{80}},
  \bibinfo{pages}{013825} (\bibinfo{year}{2009}).

\bibitem[{\citenamefont{Ono and Hofmann}(20010)}]{Ono2010}
\bibinfo{author}{\bibfnamefont{T.}~\bibnamefont{Ono}} \bibnamefont{and}
  \bibinfo{author}{\bibfnamefont{H.~F.} \bibnamefont{Hofmann}},
  \bibinfo{journal}{Phys. Rev. A} \textbf{\bibinfo{volume}{81}},
  \bibinfo{pages}{033819} (\bibinfo{year}{20010}).

\bibitem[{\citenamefont{D'Ariano et~al.}(2000)\citenamefont{D'Ariano, Paris,
  and Sacchi}}]{D'Ariano2000}
\bibinfo{author}{\bibfnamefont{G.~M.} \bibnamefont{D'Ariano}},
  \bibinfo{author}{\bibfnamefont{M.~G.~A.} \bibnamefont{Paris}},
  \bibnamefont{and} \bibinfo{author}{\bibfnamefont{M.~F.}
  \bibnamefont{Sacchi}}, \bibinfo{journal}{Phys. Rev. A}
  \textbf{\bibinfo{volume}{62}}, \bibinfo{pages}{023815}
  (\bibinfo{year}{2000}).

\bibitem[{\citenamefont{Monras}(2006)}]{Monras2006}
\bibinfo{author}{\bibfnamefont{A.}~\bibnamefont{Monras}},
  \bibinfo{journal}{Phys. Rev. A} \textbf{\bibinfo{volume}{73}},
  \bibinfo{pages}{033821} (\bibinfo{year}{2006}).

\bibitem[{\citenamefont{Olivares and Paris}(2009)}]{Olivares2009}
\bibinfo{author}{\bibfnamefont{S.}~\bibnamefont{Olivares}} \bibnamefont{and}
  \bibinfo{author}{\bibfnamefont{M.~G.~A.} \bibnamefont{Paris}},
  \bibinfo{journal}{J. Phys. B} \textbf{\bibinfo{volume}{42}},
  \bibinfo{pages}{055506} (\bibinfo{year}{2009}).

\bibitem[{\citenamefont{Gaiba and Paris}(2009)}]{Gaiba2009}
\bibinfo{author}{\bibfnamefont{R.}~\bibnamefont{Gaiba}} \bibnamefont{and}
  \bibinfo{author}{\bibfnamefont{M.~G.~A.} \bibnamefont{Paris}},
  \bibinfo{journal}{Phys. Lett. A} \textbf{\bibinfo{volume}{373}},
  \bibinfo{pages}{934} (\bibinfo{year}{2009}).

\bibitem[{\citenamefont{Higgins et~al.}(2009)\citenamefont{Higgins, Berry,
  Bartlett, Mitchell, Wiseman, and Pryde}}]{Higgins2009}
\bibinfo{author}{\bibfnamefont{B.~L.} \bibnamefont{Higgins}},
  \bibinfo{author}{\bibfnamefont{D.~W.} \bibnamefont{Berry}},
  \bibinfo{author}{\bibfnamefont{S.~D.} \bibnamefont{Bartlett}},
  \bibinfo{author}{\bibfnamefont{M.~W.} \bibnamefont{Mitchell}},
  \bibinfo{author}{\bibfnamefont{H.~M.} \bibnamefont{Wiseman}},
  \bibnamefont{and} \bibinfo{author}{\bibfnamefont{G.~J.} \bibnamefont{Pryde}},
  \bibinfo{journal}{New J. Phys.} \textbf{\bibinfo{volume}{11}},
  \bibinfo{pages}{073023} (\bibinfo{year}{2009}).

\bibitem[{\citenamefont{Cover and Thomas}(2006)}]{Cover2006}
\bibinfo{author}{\bibfnamefont{T.~M.} \bibnamefont{Cover}} \bibnamefont{and}
  \bibinfo{author}{\bibfnamefont{J.~A.} \bibnamefont{Thomas}},
  \emph{\bibinfo{title}{Elements of Information Theory}}
  (\bibinfo{publisher}{J. Wiley \& Sons, Inc.}, \bibinfo{address}{Hoboken, New
  Jersey}, \bibinfo{year}{2006}), \bibinfo{edition}{second edition} ed.

\bibitem[{\citenamefont{Hofmann}(2009)}]{Hofmann2009}
\bibinfo{author}{\bibfnamefont{H.~F.} \bibnamefont{Hofmann}},
  \bibinfo{journal}{Phys. Rev. A} \textbf{\bibinfo{volume}{79}},
  \bibinfo{pages}{033822} (\bibinfo{year}{2009}).

\bibitem[{\citenamefont{Durkin}(2010)}]{Durkin2010}
\bibinfo{author}{\bibfnamefont{G.~A.} \bibnamefont{Durkin}},
  \bibinfo{journal}{New J. Phys.} \textbf{\bibinfo{volume}{12}},
  \bibinfo{pages}{023010} (\bibinfo{year}{2010}).

\bibitem[{\citenamefont{Shannon}(1948)}]{Shannon1948}
\bibinfo{author}{\bibfnamefont{C.~E.} \bibnamefont{Shannon}},
  \bibinfo{journal}{The Bell System Technical Journal}
  \textbf{\bibinfo{volume}{27}}, \bibinfo{pages}{379} (\bibinfo{year}{1948}).

\bibitem[{\citenamefont{Bahder and Lopata}(2006{\natexlab{b}})}]{Bahder2007}
\bibinfo{author}{\bibfnamefont{T.~B.} \bibnamefont{Bahder}} \bibnamefont{and}
  \bibinfo{author}{\bibfnamefont{P.~A.} \bibnamefont{Lopata}}, in
  \emph{\bibinfo{booktitle}{The 8th International Conference on Quantum
  Communication, Measurement, and Computing}} (\bibinfo{address}{Tsukuba,
  Japan}, \bibinfo{year}{2006}{\natexlab{b}}), pp. \bibinfo{pages}{369--372},
  \urlprefix\url{http://xxx.lanl.gov/abs/quant-ph/0701243}.

\bibitem[{\citenamefont{Ou}(1997)}]{Ou1997}
\bibinfo{author}{\bibfnamefont{Z.~Y.} \bibnamefont{Ou}},
  \bibinfo{journal}{Phys. Rev. A} \textbf{\bibinfo{volume}{55}},
  \bibinfo{pages}{2598} (\bibinfo{year}{1997}).

\bibitem[{\citenamefont{Giovannetti et~al.}(2004)\citenamefont{Giovannetti,
  Lloyd, and Maccone}}]{Giovannetti2004}
\bibinfo{author}{\bibfnamefont{V.}~\bibnamefont{Giovannetti}},
  \bibinfo{author}{\bibfnamefont{S.}~\bibnamefont{Lloyd}}, \bibnamefont{and}
  \bibinfo{author}{\bibfnamefont{L.}~\bibnamefont{Maccone}},
  \bibinfo{journal}{Science} \textbf{\bibinfo{volume}{306}},
  \bibinfo{pages}{1330} (\bibinfo{year}{2004}).

\bibitem[{\citenamefont{Jaynes}(2009)}]{Jaynes2003}
\bibinfo{author}{\bibfnamefont{E.~T.} \bibnamefont{Jaynes}},
  \emph{\bibinfo{title}{Probability Theory the Logic of Science}}
  (\bibinfo{publisher}{Cambridge Press}, \bibinfo{address}{Cambridge, UK},
  \bibinfo{year}{2009}), \bibinfo{note}{sixth printing}.

\bibitem[{Note1()}]{Note1}
Note1, \bibinfo{note}{this statement applies to a balanced interferometer,
  whose path lengths satisfy Eq.~(\ref {PathLengths}).}

\bibitem[{\citenamefont{Chen and Jiang}(2007)}]{Chen2007}
\bibinfo{author}{\bibfnamefont{X.}~\bibnamefont{Chen}} \bibnamefont{and}
  \bibinfo{author}{\bibfnamefont{L.}~\bibnamefont{Jiang}}, \bibinfo{journal}{J.
  Phys. B: At. Mol. Phys.} \textbf{\bibinfo{volume}{40}}, \bibinfo{pages}{2799}
  (\bibinfo{year}{2007}).

\end{thebibliography}
%
\end{document}